# Quantization of Planetary Systems and its Dependency on Stellar Rotation

Jean-Paul A. Zoghbi[*]

**ABSTRACT**

With the discovery of now more than 500 exoplanets, we present a statistical analysis of the planetary orbital periods and their relationship to the rotation periods of their parent stars. We test whether the structure of planetary orbits, i.e. planetary angular momentum and orbital periods are 'quantized' in integer or half-integer multiples with respect to the parent stars' rotation period. The Solar System is first shown to exhibit quantized planetary orbits that correlate with the Sun's rotation period. The analysis is then expanded over 443 exoplanets to statistically validate this quantization and its association with stellar rotation. The results imply that the exoplanetary orbital periods are highly correlated with the parent star's rotation periods and follow a discrete half-integer relationship with orbital ranks $n$=0.5, 1.0, 1.5, 2.0, 2.5, etc. The probability of obtaining these results by pure chance is $p<0.024$. We discuss various mechanisms that could justify this planetary quantization, such as the hybrid gravitational instability models of planet formation, along with possible physical mechanisms such as inner discs magnetospheric truncation, tidal dissipation, and resonance trapping. In conclusion, we statistically demonstrate that a quantized orbital structure should emerge naturally from the formation processes of planetary systems and that this orbital quantization is highly dependent on the parent stars rotation periods.

## 1. INTRODUCTION

The discovery of now more than 500 exoplanets has provided the opportunity to study the various properties of planetary systems and has considerably advanced our understanding of planetary formation processes. One long suspected property of planetary systems has been the quantum-like feature that resembles the mathematical regularity of the empirical Titius-Bode (TB) law in the Solar System. Various research papers have suggested that such 'quantized' features and empirical relationships might be possible in extra-solar multi-planetary systems, such as Nottale et al. 1996, 1997a, 1997b, 2004, Rubcic & Rubcic 1998 and 1999, Poveda & Lara 2008, and Chang 2010, just to mention a few. In case they truly exist, one main question that needs to be answered is what physical processes might cause these 'quantization' features to develop. The gravitational instability model of planet formation has been successfully used in the past to explain 'discrete' power law distributions in planetary spacing (Griv & Gedalin 2005). Similarly, hybrid models of planetary formation (e.g. Durisen et al. 2005), are characterized by concentric dense gas rings that are produced by resonances and discrete spiral modes  which, in theory, can be correlated to orbital 'quantization' features. Similarly, tidal dissipation and the role of angular momentum transfer along with mean-motion resonances and resonance trapping, play an important role in the final orbital configuration. In all of these mechanisms, the stellar rotation period is a critical parameter. The main motivation in this paper is therefore to (a) statistically search for any apparent quantum-like features in the orbital structure of exoplanetary systems, (b) to determine whether the quantization parameters are related to any specific physical system property (the stellar rotation rate is examined in this paper), and (c) to shed some light on the nature of the possible physical processes that might lead to this apparent quantization. We will argue on dynamical terms that a quasi-quantum model might emerge naturally from the formation processes that determine the final configuration of a planetary system.

The plan of this paper is as follows: Sect. 2 describes the basic methodology and simple quantum-like model. In Sect. 3, the model is applied to the Solar System. In Sect. 4, the analysis is expanded over a sample of 443 exoplanets, for which we could obtain stellar rotation periods. In Sect. 5, a statistical analysis of the results is presented demonstrating that the specific angular momenta of all planetary orbits generally follow half-integer multiples of the specific angular momentum at the parent star's corotation radius. Sect. 6 briefly proposes various physical mechanisms that may justify the obtained results. Prospects and conclusions are drawn in Sect. 7.

## 2. METHODOLGY

We will be testing for the quantization of planetary angular momentum, i.e. we will test that planetary angular momenta ought to have discrete values in multiples of a 'ground-state' system-specific parameter. Within an order of magnitude estimate, the corotation orbit represents an approximate inferior limit to the position of planetary orbits. This is confirmed by various physical mechanisms that are discussed in Section 6, such as spin-orbit coupling and tidal dissipation, as well as disc-locking and magnetic braking which create a barrier and inferior limit to planetary migration. On that basis, we postulate that the corotation orbit represents the ground-state orbit of planetary systems and assign to it the orbital rank $n$=1. However, this does not negate the possibility of having physical objects orbiting inside the corotation orbit. Nevertheless, the corotation orbit (at $n$=1) is particularly chosen because of its importance as a base reference to the orbital parameters of the entire planetary system, and in particular, their relationship to the parent star's rotation period.

The corotation radius $r_0$ is defined in terms of the star's rotation rate $\Omega_s$ by

$$r_0 = \left( \frac{GM}{\Omega_s{}^2} \right)^{1/3}, \qquad (2.1)$$

where $G$ is the gravitational constant, $M$ and $\Omega_s$ are the mass and rotation rate of the parent star respectively. Similarly, the mean motion orbital velocity $v_0$ and specific angular momentum $J_0$ (per unit mass) at the corotation radius are given by

$$v_0 = \left( \frac{GM}{r_0} \right)^{1/2} = \left( GM\Omega_s \right)^{1/3}, \qquad (2.2)$$

$$J_0 = v_0 r_0 = \left( \frac{G^2 M^2}{\Omega_s} \right)^{1/3} \qquad (2.3)$$

If our planetary quantization hypothesis is valid, then the specific angular momentum $J_n$ of any planetary orbit $n$ would follow discrete and quantized multiples of the specific angular momentum $J_0$ at the corotation orbit $n$=1.

$$J_n = nJ_0 = n \left( \frac{G^2 M^2}{\Omega_s} \right)^{1/3} \qquad (2.4)$$

In other words, the ratio of the specific orbital angular momentum $J_n$ of a planet in the $n^{th}$ planetary orbit to the specific angular momentum $J_0$ at the 'ground-state' corotation orbit ought to be incremental by a discrete value.

$$n = \frac{J_n}{J_0} \qquad\qquad (2.5)$$

For nearly circular orbits, Newton's force balance equation of motion gives,

$$r_n = \frac{GM}{v_n^2} \text{ , and} \qquad\qquad (2.6)$$

$$J_n = v_n r_n = (GMr_n)^{\frac{1}{2}} \qquad\qquad (2.7)$$

Combining equation (2.1), (2.5), and (2.7), we obtain an $n^2$ law for the quantized semi-major axis $r_n$ of the $n^{th}$ orbit, in terms of the corotation radius $r_0$, spin rotation rate $\Omega_s$ of the central star, and orbital rank $n$ by

$$r_n = n^2 r_0 = n^2 \left[ \frac{GM}{\Omega_s^2} \right]^{\frac{1}{3}} \qquad\qquad (2.8)$$

From Kepler's Third Law and eq. (2.8), the quantized orbital period $P_n$ of the $n^{th}$ planetary orbit is also given in terms of the corotation orbital period $P_0$ and orbital rank $n$ by

$$P_n = n^3 P_0 = n^3 P_{rot} \text{ , or } n = \left( \frac{P_n}{P_{rot}} \right)^{\frac{1}{3}} \qquad\qquad (2.9)$$

Where $P_n$ is the planet's orbital period and $P_0$ is the corotation period which is by definition equal to $P_{rot}$, the star's rotation period.

## 3. SOLAR SYSTEM APPLICATION & RESULTS

### 3.1 The Solar System Orbital Ranks

The quantum-like model described in Sect. 2 is first applied to the Solar System in order to discern any discrete pattern in their orbital ranks $n$. We will calculate the orbital ranks $n$ using the Sun's rotation period $P_{Sun}$ taken from *Allan's Astrophysical Quantities* (Cox 1999) as 25.38 days and the solar rotation rate $\Omega_{Sun}$=2.8 x $10^{-6}$ rd $s^{-1}$. The Sun's corotation specific angular momentum $J_0$ is calculated from Eq. (2.3) and will represent the base quantization parameter for all possible planetary orbits in the Solar System. The planetary orbital ranks are first calculated using the Sun's present rotation period (25.38 days) and presented in Table 1. However, since the Sun's rotation rate has already decayed with age through angular momentum loss, the orbital ranks are also calculated using the Sun's rotation rate at the early stage of planets formation. The orbital parameters of the solar system are assumed to have settled into a long-term stable configuration at around 650 Myr or so. Using data on solar-type stars in the Hyades (age ~650 Myr), we selected star VB-15 which has a B-V index similar to the Sun to estimate the Sun's rotation period of 8 days (Radick et al. 1987) at the planets' formation age, i.e. the time when the solar system planets orbits have stabilized and the proposed quantization 'frozen-in'. The planet's orbital ranks $n$ are calculated from Eq. (2.5) for each planetary orbit using both the Sun's present and earlier formation rotation periods ($J_0$=1.8319 and 1.2467 $m^2$ $s^{-1}$ respectively). The results are presented in Table 1.

Besides the main planets, the list includes main mass distribution peaks such as the Asteroid Belt families: Flora, Ceres, Pallas, Cybele, and Thule, as well as Centaurs, trans-Neptunian Cubewanos in the Kuiper Belt, and the recently discovered Scattered Disc Object (SDO) 2003-UB313, previously dubbed as the "tenth planet".

**Table 1 -** Solar System orbital parameters [1, 2, 3] and orbital ranks $n$, calculated from the ratio of the planets' specific orbital angular momenta to that of the Solar System's corotation orbit along with the deviations $\Delta n$ from half-integer values.

| Planet/ Object | Planetary Parameters | | | Present Age (P$_0$=25.38 days) | | Formation Age (650 Myr) P$_0$=8 days | |
|---|---|---|---|---|---|---|---|
| | $r_n$ Semi-Major Axis (AU) | $e$ Orbital Eccentricity | $J_n$ Specific Angular Momentum ($10^{15}$ m$^2$ s$^{-1}$) | $n$ Orbital Rank ($J_0$=1.8319) | Deviation $\Delta n$ | $n$ Orbital Rank ($J_0$=1.2467) | Deviation $\Delta n$ |
| Corotation Orbit | 0.1690 | - | 1.8319 | 1.00 | 0.00 | 1.00 | 0.00 |
| Mercury | 0.3871 | 0.2056 | 2.7131 | 1.48 | -0.02 | 2.18 | 0.18 |
| Venus | 0.7233 | 0.0068 | 3.7896 | 2.07 | 0.07 | 3.04 | 0.04 |
| Earth | 1.0000 | 0.0167 | 4.4553 | 2.43 | -0.07 | 3.57 | 0.07 |
| Mars | 1.5237 | 0.0934 | 5.4762 | 2.99 | -0.01 | 4.39 | -0.11 |
| Flora Family | 2.2020 | 0.1561 | 6.5312 | 3.57 | 0.07 | 5.24 | -0.26 |
| Ceres Family | 2.7660 | 0.0800 | 7.3870 | 4.03 | 0.03 | 5.93 | -0.07 |
| Cybele Family | 3.4360 | 0.1040 | 8.2149 | 4.48 | -0.02 | 6.59 | 0.09 |
| Thule & Comets[a] | 4.2770 | 0.0120 | 9.2146 | 5.03 | 0.03 | 7.39 | -0.11 |
| Jupiter | 5.2034 | 0.0484 | 10.1525 | 5.54 | 0.04 | 8.14 | 0.14 |
| Comets[b] | 5.9860 | 0.0441 | 10.8914 | 5.95 | -0.05 | 8.74 | 0.24 |
| Saturn | 9.5371 | 0.0542 | 13.7407 | 7.50 | 0.00 | 11.02 | 0.02 |
| Chiron | 13.7035 | 0.3831 | 15.2365 | 8.32 | -0.18 | 12.22 | 0.22 |
| Chariklo | 15.8700 | 0.1758 | 17.4747 | 9.54 | 0.04 | 14.02 | 0.02 |
| Uranus | 19.1913 | 0.0472 | 19.4987 | 10.64 | 0.14 | 15.64 | 0.14 |
| Pholus | 20.4310 | 0.5730 | 16.5067 | 9.01 | 0.01 | 13.24 | 0.24 |
| Neptune | 30.0690 | 0.0086 | 24.4333 | 13.34 | -0.16 | 19.60 | 0.10 |
| Pluto | 39.4817 | 0.2488 | 27.1181 | 14.80 | -0.20 | 21.75 | -0.25 |
| Cubewanos[c] | 43.4050 | 0.0654 | 29.2939 | 15.99 | -0.01 | 23.50 | 0.00 |
| Eris UB313 | 67.6681 | 0.4418 | 32.8840 | 17.95 | -0.05 | 26.38 | -0.12 |

**Notes**
[a] Comets include Lovas, Denining-Fuyikawa, Kearns-Kwee, Coma-Sola , Maury, & Whipple
[b] Comets include Schwassmann-Wachmann 1 and 66/P Du-Toit 1
[c] Cubewanos include 1992QB1, Varuna, and Quaoar

From Table 1, it can be observed from the orbital ranks calculated using the Sun's present rotation period that the Solar System exhibits a discrete and quantized orbital structure where the planets' specific orbital angular momenta $J_n$ are ranked in discrete half-integer multiples of the specific angular momentum $J_0$ at the solar corotation orbit ($n$= 1.0, 1.5, 2.0, 2.5, 3.0, 3.5, etc.). The $\Delta n$ deviations from integer or half-integer values are included in Table 3 and indicate that 16 out of 19 planetary orbits have absolute deviations $|\Delta n|$<0.07. The discrete nature of planetary semi-major axes, mean orbital velocities, and orbital periods, in terms of half-integer values, follow logically from the quantized orbital angular momentum results.

The inner planets Mercury ($n$=1.5), Venus ($n$=2.0), Earth ($n$=2.5), and Mars ($n$=3.0) occupy the ranks $n$= 1.48, 2.07, 2.43, and 2.99 respectively with minimal deviations $\Delta n$ from the closest integer or half-integer values. In the main Asteroid Belt, the orbits of the Flora family are ranked at $n$=3.5, with both Flora and Ariadne occupying $n$=3.57. At the orbital rank $n$=4, the main asteroid families of Ceres and Pallas represent the group and both occupy the rank $n$=4.03. This orbital rank also includes Misa, Eunomia, Lamberta, and the Chloris families at $n$=3.90, Ino and Adeana at $n$=3.94, Dora at $n$=3.96, Elpis, Herculina, Gyptis, Juewa, Minerva, Thisbe, Dynamene, and Eunike are all at $n$= 3.99, Eugenia and Nemesis at $n$=4.0, the Lydia, Gefion, and Pompeja at $n$=4.01, and the Brasilia & Karin families at $n$=4.09.

The orbits of the Cybele family of asteroids are ranked at $n$=4.5, with the main asteroid Cybele for instance, occupying rank $n$=4.48, Sibylla and Hermione at $n$= 4.47, Bertholda at $n$= 4.49, Camilla at $n$=4.52, and Sylvia at $n$=4.53. At the next orbital rank of $n$=5, the main asteroid Thule occupies $n$=5.03. At the orbital rank of $n$=5.5, Jupiter occupies the rank $n$=5.54 along with the Trojan asteroids such as Achilles at $n$=5.48, Diomedes at $n$=5.49, Aneas at $n$=5.50, Patroclus and Nestor at $n$=5.51, Hektor at $n$=5.53, and Agamemmon at $n$=5.54. Beyond Jupiter, the orbital ranks at $n$=6.0, $n$=6.5, and $n$=7.0 do not appear to be occupied by any major object. However, this does not exclude various periodic comets whose orbital properties match several orbital ranks in the Solar System. To mention a few, the comets 29/P Schwassmann-Wachmann-1 and 66/P Du-Toit both occupy $n$=5.95 and $n$=5.94 respectively. Ranked also with the asteroid Thule for instance, are the comets 36/P Whipple at $n$=4.97, 115/P Maury at $n$=5.01, 32/P Coma-Sola at $n$=5.02, 59/P Kearns-Kwee at $n$=5.05, 72/P Denning-Fuyikawa at $n$=5.06, and 93/P Lovas at $n$=5.08. However, the unoccupied ranks beyond Jupiter are better explained by orbital migration and the outward expansion of the Solar System boundaries.

Saturn occupies the rank $n$=7.50 exactly, while 'centaurs' such as Chiron is at $n$=9.01, Chariklo is at $n$=9.69, and Pholus is at $n$=10.97. Uranus occupies the rank $n$=10.65 and Neptune $n$=13.34, with relatively higher but nearly equal and opposite deviations from integer or half-integer ranking, $\Delta n$=+0.14 & -0.16 respectively. At the orbital rank of $n$=15.5, Pluto occupies $n$=15.28. Beyond Pluto and at the orbital rank $n$=16, some Cubewanos, classified as trans-Neptunian objects (TNO) in the Kuiper Belt are included, such as Quaoar at $n$=16.00 and Varuna at $n$=15.97. The recently discovered Scattered Disc Object (SDO) Eris UB313 occupies the rank $n$=17.95. This can be used to predict the location of objects within and beyond SDOs. At $n$=20 for instance, an object may be discovered orbiting at 67.85 AU.

We note that the deviations $\Delta n$ from the closest integer or half-integer are negligible up to Saturn and all orbital ranks are occupied by planets or asteroid mass peaks within that region. Beyond Saturn's orbit, the deviations $\Delta n$ are relatively higher for Uranus, Neptune and Pluto and several orbital ranks are vacant. One possible explanation may be related to dissipation in the solar protoplanetary disc that allows both inward and outward planetary migrations, depending on the initial position and the radius of maximum viscous stress located just outside the orbit of Saturn (at around 10 AU). Hence, the orbits of protoplanets forming within that critical radius tend to compact, while those forming outside that radius are stretched outwards. This could explain the relatively higher $\Delta n$ deviations beyond Saturn's orbit and, more importantly, the unoccupied orbital ranks produced by the outward expansion.

The orbital ranks $n$ that were calculated using the Sun's rotation period (8 days) at the formation age of 650 Myr also exhibit a discrete and quantized structure, albeit with higher deviations $\Delta n$ from half-integer numbers. It can be therefore inferred that the decay in solar rotation rate has improved or at least had a minimal effect on the quantized orbital ranks, most likely because the orbital ranks are proportional to the cubic root (1/3) of the decreasing rotation rate (Eq. 2.5). The slowing down of the Sun's rotation rate has actually improved the discrete quantized nature of the orbital structure with deviations from half-integers approaching zero as the rotation rate decreases asymptotically with age (Skumanich 1972), where it reaches a limit value that has negligible further effect to the orbital rank values. This effect is more clearly seen in Section 4.1 where the model is applied to 443 exoplanets.

## 4. EXOPLANETARY APPLICATION & RESULTS

### 4.1 Exoplanetary Orbital Ranks at parent stars' present Rotation Periods

To date, more than 500 exoplanets with 49 multi-planetary systems have been discovered. In order to verify whether the quantization of planetary angular momentum in discrete half-integer values is a universal occurrence and not just a coincidence of the Solar System, and in order to validate the dependency of this quantization on stellar rotation, a sample of 443 exoplanets, for which star rotation data is available, is analyzed with respect to the rotation periods of their parent stars. Out of the 443 exoplanets, almost half (216 stars) have host stars with rotation periods available from literature or measured from log R'$_{H K}$. These were obtained from the planets' discovery papers (49 parent stars see references in Table 3) or from Watson et al. 2010 (167 parent stars) which conveniently compiles all published rotation periods of exoplanetary host stars in Table 1 of that paper. We preferred to use Table 1 of Watson et al. 2010 and not Tables 2 & 3, as Table 1 compiles published rotation periods while Table 2 & 3 use Markov-Chain Monte Carlo simulation to estimate them. Out of the 167 rotation periods in Watson et al. 2010, 7 stars have actual observed rotation period and these are: rho CrB (17 days), Tau Boo (14 days), Epsilon Eri (11.68 days), HD 3651 (44 days), HD 62509 (135 days), HD 70573 (3.3 days), HD 89744 (9 days) (Watson et al. 2010). The remaining 227 stellar rotation periods were estimated from the projected rotational velocities *v*sin*i* and stellar radii, with certain levels of uncertainty. We have used the 'Catalogue of Nearby Exoplanets' (Butler et al. 2006), the 'Catalogue of Rotational Velocities' (Glebocki et al. 2005), the Exoplanet Data Explorer Table http://exoplanets.org/ Wright & Marcy 2010, along with some planet discovery papers to obtain values of *v*sin*i* (see references in Table 3). We noted that in many cases the values of *v*sin*i* listed in the Exoplanet Data Explorer Table were truncated and rounded up, which is why we attempted as much as possible to obtain more accurate values (to 2 significant digits) from the referenced papers. Moreover, in some cases where the catalogues listed multiple values of *v*sin*i* for a particular star, the values that are listed as upper limits were generally avoided and similar values when measured and corroborated by different sources were selected. The absolute stellar radii were mainly taken from the referenced planet discovery papers, the Fundamental Parameters of Stars Catalogue (Allende Prieto & Lambert 1999), the Catalogue of Stellar Diameters (Pasinetti-Fracassini et al. 2001), or the Effective Temperatures and Radii of Stars Catalogue (Masana et al 2006).

### 4.2 Uncertainty Considerations

The vast majority of the *v*sin*i* values we found are under 4 km/sec. The inherent measurement uncertainty in these *v*sin*i* values is at best around 0.5 km/sec and at worst 1 to 2 km/sec. As for stellar radii, although they can be determined to a precision of the order of 5 percent for the small minority of planets that transit their parent stars, the radius estimates for the remainder are unlikely to be established to a precision better than 10 percent. Moreover, the orbital angular momentum for exoplanets depend on stellar mass which are mostly derived from isochrones fits and have an inherent uncertainty in the order 10 percent. Therefore, the extra-solar orbital ranks if calculated from the planetary angular momenta in eq. (2.5) would thus be uncertain by at least 30 percent (since M$_s$ and R$_s$ are correlated on the main sequence) even before the uncertainties in the measured *v*sin*i* (a further 30 percent) and the effects of unknown orbital inclination are taken into account.

Fortunately the dependence on stellar mass in Eq. (2.5) can be eliminated. Equation (2.9) is the key, as it uses the orbital period of the planet and rotation period of the star directly. By using Eq. (2.9) to calculate the orbital ranks, the use of the planet's semi-major axis is avoided, as it requires knowledge of the uncertain stellar mass. Since the orbital period is one of the few planetary parameters that is measured directly and with high precision, this is the best quantity to use. The same is true for almost half of the exoplanets sample having directly measure stellar rotation periods. The remaining rotation periods were derived from a measurement of *v*sin*i* which are typically uncertain by 20 to 30 percent and estimates of stellar radii are also uncertain by roughly 10 percent. This may not be disastrous, since the ratio of the two values is raised to the one-third power. Therefore, the final estimates for the orbital ranks *n* are expected to be uncertain by 5 to 10 percent. Moreover, the uncertainty due the unknown inclination of the stellar rotation axis takes the form of $(\sin i)^{-1/3}$. For inclination angles ranging from 45° to 90°, this factor is very close to unity, and therefore has an insignificant effect on the calculated orbital ranks *n*. For inclinations between 30° to 45°, the $(\sin i)^{-1/3}$ factor can affect the *n* values by as much as 10 to 20 per cent. However, since the most likely inclination of a random stellar sample is 57° (Trilling et al. 2002) and because the radial velocity technique is biased towards detecting planetary systems with inclinations near 90°, the average value of $\sin i$ is expected to range between $\pi/4$ and unity. With the number of exoplanets under consideration, the average value of $(\sin i)$ for the population approaches the value for a random distribution. Hence, the most likely effect of the inclination factor on the calculated orbital ranks should again not exceed 7 to 10 per cent on average. A Monte-Carlo treatment is used in Sect. 5 to study the effect of these uncertainties.

In the first approach, the extra-solar orbital ranks *n* and their related deviations from half-integer values $\Delta n$ are calculated using the ratio of the planet's orbital period to that of the parent star's current rotation period, as in Eq. (2.9) and presented in Table 3.

### 4.3 Exoplanetary Orbital Ranks at the Planetary Formation Epoch (~650 Myr)

In the second approach, we address  the concern that the half-integer orbital ranks are calculated using the rotation period for the present age of the star, and not at the epoch when planetary systems were formed, when it is known that solar-type stars observed in young star clusters do not rotate at constant rates throughout their lifetimes. We therefore need to study whether any quantization feature exist at the formation age when supposedly it gets 'frozen in'.

The rotation rates of stars with outer convection zones generally decay with age, approximately as the inverse square root of time (Skumanich 1972) through angular momentum loss via hot magnetically-channeled winds. However, at around 600 Myr or so, planetary systems eventually settle into long-term stable configurations and their orbital periods are constant while the stellar rotation periods continue to increase. However, Soderblom et al. (2001) indicated that the rotation of solar-type stars, in evolving from the Pleiades (100 Myr) to the Hyades (650 Myr), changes only modestly in the mean, but undergoes a huge convergence in the spread of rotation rates. Thus, at any one mass in the Pleiades (100 Myr), the range of rotation rates vary by an order of magnitude or more, yet in the Hyades (650 Myr), stars of the same mass have nearly identical rotation rates. The convergence occurs for upper bound rotation rates, as the lower bounds of both clusters are nearly identical (Soderblom et al. 2001). Since most of our sample exoplanetary stars have rotation periods in the lower bounds, we can assume that these have remained essentially unchanged over the period 100 - 650 Myr (Soderblom et al. 2001), i.e. during the period of planetary formation. With this minimal decay in rotation periods, it is therefore logical to expect a quantized distribution of orbital ranks around half-integer values at the early formation age period.

Nevertheless, rather than using only the present rotation periods of these stars, we additionally adopt a stellar rotation period at the fiducial planets formation age (650 Myr), that can be derived from the star's B-V color and the known rotation periods of stars of the same color in the Hyades (aged 650 Myr). This procedure has the advantage that it can be based on direct measurements of stellar rotation periods in stellar clusters of known age, near the epoch when the proposed planetary quantization would have been established. This bypasses the difficulties arising from the uncertainties in *v*sin*i*, inclination, stellar radius, and stellar age.

Table 2 presents the Hyades stars, their B-V color, and directly measured rotation periods, which were used in matching the 443 exoplanetary star sample. The Hyades stars' B-V color range from 0.41 to 1.53. However, we could not find any measured rotation periods for Hyades stars in the B-V range between 0.30 to 0.40, 0.69 to 0.73, 0.78 to 0.80, and 0.87 to 0.88. Instead, five Hyades stars (HD 28911, HD 26756, HD 27282, HD 21663, and HD 26397) with rotation periods derived from *v*sin*i* and stellar radii were selected to supplement for the missing B-V ranges. Additionally, one Praesepe star H218 (of similar age ~ 650 Myr), having directly measured rotational period, was selected to cover for the few exoplanetary stars of similar spectral type M2 to M4 (Scholz & Eisloffel 2002). A Monte-Carlo treatment is presented in Sect. 5 to address the inherent uncertainty in these rotation periods.

The B-V color values for the exoplanetary stars were obtained primarily from the 'All-sky Compiled Catalogue of 2.5 million stars' (Kharchenko 2001), the 'NOMAD Catalog' (Zacharias et al. 2005), and the 'Hipparcos & Tycho Catalogue' (ESA 1997) and were matched with the corresponding value from the Hyades stars to obtain an estimate of their early rotation periods at the age of 650 Myr. These rotation periods were then used to calculate the orbital ranks *n* near the planet formation epoch when the proposed quantization would have been established.

Table 2 - Hyades Stars Rotation Periods used to match the B-V color of exoplanetary stars

| VB No. | B-V | Represents B-V Range | $P_{rot}$ (days) | Ref. | VB No. | B-V | Represents B-V Range | $P_{rot}$ (days) | Ref. |
|--------|------|---------------------|------|------|--------|------|---------------------|-------|------|
| VB 94 | 0.396 | Less than 0.430 | 1.67 | (2) | VB 17 | 0.706 | 0.690 to 0.710 | 7.25 | (2) |
| VB 78 | 0.451 | 0.431 to 0.460 | 2.90 | (1) | VB 27 | 0.721 | 0.711 to 0.730 | 7.15 | (2) |
| VB 81 | 0.470 | 0.460 to 0.480 | 2.80 | (1) | VB 92 | 0.736 | 0.731 to 0.740 | 9.13 | (6) |
| VB 121 | 0.500 | 0.480 to 0.510 | 3.70 | (7) | VB 26 | 0.745 | 0.741 to 0.760 | 9.06 | (5) |
| VB 48 | 0.518 | 0.511 to 0.520 | 2.50 | (1) | VB 22 | 0.770 | 0.761 to 0.780 | 5.61 | (7) |
| VB 65 | 0.535 | 0.521 to 0.530 | 5.87 | (5) | VB 3 | 0.786 | 0.781 to 0.799 | 12.04 | (2) |
| VB 59 | 0.543 | 0.531 to 0.545 | 5.13 | (5) | VB 21 | 0.816 | 0.800 to 0.819 | 5.49 | (3) |
| VB 29 | 0.548 | 0.546 to 0.559 | 3.00 | (1) | VB 79 | 0.831 | 0.820 to 0.839 | 9.71 | (6) |
| VB 119 | 0.563 | 0.560 to 0.569 | 4.00 | (7) | VB 153 | 0.855 | 0.840 to 0.869 | 9.18 | (4) |
| VB 31 | 0.572 | 0.570 to 0.579 | 4.72 | (3) | VB 138 | 0.871 | 0.870 to 0.879 | 19.19 | (2) |
| VB 52 | 0.592 | 0.580 to 0.599 | 5.64 | (3) | VB 43 | 0.907 | 0.880 to 0.920 | 10.26 | (5) |
| VB 50 | 0.604 | 0.600 to 0.609 | 5.10 | (1) | VB 91 | 0.936 | 0.921 to 0.970 | 9.36 | (5) |
| VB 73 | 0.609 | 0.610 to 0.620 | 7.38 | (6) | VB 25 | 0.984 | 0.971 to 1.010 | 12.64 | (5) |
| VB 97 | 0.624 | 0.621 to 0.629 | 6.45 | (3) | VB 175 | 1.031 | 1.011 to 1.080 | 10.82 | (5) |
| VB 18 | 0.640 | 0.630 to 0.649 | 8.65 | (4) | VB 181 | 1.167 | 1.081 to 1.199 | 11.92 | (5) |
| VB 63 | 0.651 | 0.650 to 0.653 | 7.73 | (5) | VB 173 | 1.237 | 1.200 to 1.300 | 14.14 | (5) |
| VB 15 | 0.657 | 0.654 to 0.660 | 7.43 | (3) | VB 190 | 1.357 | 1.301 to 1.499 | 3.66 | (5) |
| VB 64 | 0.664 | 0.661 to 0.670 | 8.64 | (6) | H218 | M3 M4 | 1.511 to 1.600 | 0.68 | (8) |
| VB 58 | 0.680 | 0.671 to 0.689 | 6.20 | (1) | | | | | |

**References for Hyades Rotational Periods:** (1) Duncan et al. 1984, (2) Glebocki et al. 2000, (3) Paulsen et al. 2003, (4) Paulsen et al. 2004, (5) Radick et al. 1987, (6) Radick et al. 1995, (7) Rutten et al. 1987, (8) Scholz & Eisloffel 2007

In Table 3, the orbital rank for each exoplanet is calculated using both the present stellar rotation period and the fiducial star rotation period at formation age (650 Myr) and the following is presented:

a. The exoplanetary orbital periods, obtained from the updated online database of the Exoplanets Encyclopedia website (Schneider 2010)

b. The present orbital ranks *n* and deviations from half-integer Δ*n*, calculated from the current stellar rotation periods at present, either directly measured $P_{rot}$ or derived from *v*sin*i* and the stellar radii (see *v*sin*i* and $P_{rot}$ references).

c. The fiducial orbital ranks *n* at formation age (650 Myr) and deviations from half-integer Δ*n*, calculated from directly-measured rotation periods at the fiducial age of 650 Myr from Hyades stars having B-V values that match those of the exoplanetary stars (Table 2 above).

**Table 3** – Orbital ranks *n* for the 443 exoplanets calculated from their planetary orbital period $P_n$ and the parent star's rotation period $P_{rot}$ (days), both at present and at the fiducial age of 650 Myr

| Exoplanet | Planet Orbital Period $P_n$ | Stellar Radius (Rs) | *v* sin *i* (km s⁻¹) | Present Rotation Period $P_{rot}$ | Present Orbital Ranks *n* | Δ*n* | Ref. *vsini* & $P_{rot}$ | B-V | Rotation Period $P_{rot}$ at 650 Myr | Orbital Ranks *n* at 650 Myr | Δ*n* |
|---|---|---|---|---|---|---|---|---|---|---|---|
| 11 UMi b | 516.22 | 24.08 | 1.50 | 811.96 | 0.86 | -0.14 | (75) | 1.391 | 3.66 | 5.21 | 0.21 |
| 14 And b | 185.84 | 11.00 | 2.60 | 213.99 | 0.95 | -0.05 | (75) | 1.029 | 10.82 | 2.58 | 0.08 |
| 14 Her b | 1,773.40 | | | 41.00 | 3.51 | 0.01 | (73) | 0.884 | 10.26 | 5.57 | 0.07 |
| 16 Cyg Bb | 799.50 | | | 29.34 | 3.01 | 0.01 | (73) | 0.663 | 8.64 | 4.52 | 0.03 |
| 24 Sex b | 452.80 | 5.13 | 2.77 | 93.65 | 1.69 | 0.19 | (40) | 0.920 | 10.26 | 3.53 | 0.03 |
| 24 Sex c | 883.00 | 4.90 | 2.77 | 89.47 | 2.14 | 0.14 | (40) | 0.920 | 10.26 | 4.42 | -0.08 |
| 30 Ari B b | 335.10 | 1.13 | 38.50 | 1.48 | 6.09 | 0.09 | (26) | 0.410 | 1.67 | 5.85 | -0.15 |
| 4 Uma b | 269.30 | 18.11 | 1.00 | 915.98 | 0.66 | 0.16 | (24) | 1.198 | 14.40 | 2.65 | 0.15 |
| 47 Uma b | 1,078 | | | 23.00 | 3.61 | 0.11 | (73) | 0.624 | 6.45 | 5.51 | 0.03 |
| 47 Uma c | 2,391 | | | 23.00 | 4.70 | 0.20 | (73) | 0.624 | 6.45 | 7.18 | 0.18 |
| 47 Uma d | 14,002 | | | 23.00 | 8.48 | -0.02 | (73) | 0.624 | 6.45 | 12.95 | -0.05 |
| 51 Peg b | 4.23 | | | 29.50 | 0.52 | 0.02 | (73) | 0.673 | 6.20 | 0.88 | -0.12 |
| 55 Cnc b | 14.65 | | | 42.20 | 0.70 | 0.20 | (34) | 0.870 | 19.19 | 0.91 | -0.09 |
| 55 Cnc c | 44.34 | | | 42.20 | 1.02 | 0.02 | (34) | 0.870 | 19.19 | 1.32 | -0.18 |
| 55 Cnc d | 5,218 | | | 42.20 | 4.98 | -0.02 | (34) | 0.870 | 19.19 | 6.48 | -0.02 |
| 55 Cnc e | 2.82 | | | 42.20 | 0.41 | -0.09 | (34) | 0.870 | 19.19 | 0.53 | 0.03 |
| 56 Cnc f | 260.00 | | | 42.20 | 1.83 | -0.17 | (34) | 0.870 | 19.19 | 2.38 | -0.12 |
| 6 Lyn b | 874.77 | 5.20 | 1.32 | 199.25 | 1.64 | 0.14 | (75) | 0.934 | 10.26 | 4.40 | -0.10 |
| 61 Vir b | 4.22 | | | 29.00 | 0.53 | 0.03 | (72) | 0.709 | 7.25 | 0.83 | -0.17 |
| 61 Vir c | 38.02 | | | 29.00 | 1.09 | 0.09 | (72) | 0.709 | 7.25 | 1.74 | 0.24 |
| 61 Vir d | 123.01 | | | 29.00 | 1.62 | 0.12 | (72) | 0.709 | 7.25 | 2.57 | 0.07 |
| 70 Vir b | 116.69 | | | 35.80 | 1.48 | -0.02 | (73) | 0.710 | 7.25 | 2.52 | 0.02 |
| 81 Cet b | 952.70 | 11.00 | 1.80 | 309.09 | 1.46 | -0.04 | (75) | 1.021 | 10.82 | 4.45 | -0.05 |
| BD +14 4559b | 268.94 | 0.95 | 2.50 | 19.22 | 2.41 | -0.09 | (75) | 0.980 | 12.64 | 2.77 | -0.23 |
| BD +20 2457b | 379.63 | | | 2460 | 0.54 | 0.04 | (76) | 1.250 | 14.14 | 2.99 | -0.01 |
| BD +20 2457c | 621.99 | | | 2460 | 0.63 | 0.13 | (76) | 1.250 | 14.14 | 3.53 | 0.03 |
| BD -08 2823b | 5.60 | | | 26.60 | 0.59 | 0.09 | (30) | 1.071 | 10.82 | 0.80 | -0.20 |
| BD -08 2823c | 237.60 | | | 26.60 | 2.07 | 0.07 | (30) | 1.071 | 10.82 | 2.80 | -0.20 |
| BD -17 63 b | 655.60 | | | 39.00 | 2.56 | 0.06 | (51) | 1.128 | 11.92 | 3.80 | -0.20 |
| BD-10 3166 b | 3.49 | 1.71 | 0.90 | 96.10 | 0.33 | -0.17 | (75) | 0.850 | 9.18 | 0.72 | 0.22 |
| CoRoT-1 b | 1.51 | 1.11 | 5.20 | 10.80 | 0.52 | 0.02 | (8) | 0.280 | 1.67 | 0.97 | -0.03 |
| CoRoT-10 b | 13.24 | 0.79 | 2.00 | 19.98 | 0.87 | -0.13 | (11) | 1.460 | 3.66 | 1.54 | 0.04 |
| CoRoT-11 b | 2.99 | 1.37 | 40.00 | 1.73 | 1.20 | 0.20 | (22) | 0.930 | 9.36 | 0.68 | 0.18 |
| CoRoT-12 b | 2.83 | 1.12 | 2.90 | 19.46 | 0.53 | 0.03 | (23) | 0.828 | 9.71 | 0.66 | 0.16 |
| CoRoT-13 b | 4.04 | 1.01 | 4.00 | 12.77 | 0.68 | 0.18 | (18) | 0.807 | 5.49 | 0.90 | -0.10 |
| CoRoT-14 b | 1.51 | | | 5.70 | 0.64 | 0.14 | (68) | | | | |
| CoRoT-2 b | 1.74 | | | 4.55 | 0.73 | 0.23 | (1) | 0.237 | 1.67 | 1.01 | 0.01 |
| CoRoT-3 b | 4.26 | 1.56 | 17.00 | 4.64 | 0.97 | -0.03 | (75) | 0.538 | 5.13 | 0.94 | -0.06 |
| CoRoT-4 b | 9.20 | 1.15 | 6.40 | 9.09 | 1.00 | 0.00 | (75) | 0.099 | 1.67 | 1.77 | -0.23 |
| CoRoT-5 b | 4.04 | 1.19 | 1.00 | 59.99 | 0.41 | -0.09 | (75) | 0.670 | 8.64 | 0.78 | -0.22 |
| CoRoT-6 b | 8.89 | 1.03 | 7.50 | 6.91 | 1.09 | 0.09 | (75) | 0.330 | 1.67 | 1.75 | 0.25 |
| CoRoT-7 b | 0.85 | | | 23.00 | 0.33 | -0.17 | (77) | 0.856 | 9.18 | 0.45 | -0.05 |

| Exoplanet | Planet Orbital Period $P_n$ | Stellar Radius (Rs) | $v \sin i$ (km s⁻¹) | Present Rotation Period $P_{rot}$ | Present Orbital Ranks $n$ | $\Delta n$ | Ref. $v\sin i$ & $P_{rot}$ | B-V | Rotation Period $P_{rot}$ at 650 Myr | Orbital Ranks $n$ at 650 Myr | $\Delta n$ |
|---|---|---|---|---|---|---|---|---|---|---|---|
| CoRoT-7 c | 3.70 | | | 23.00 | 0.54 | 0.04 | (77) | 0.856 | 9.18 | 0.74 | 0.24 |
| CoRoT-8 b | 6.21 | 0.77 | 2.00 | 19.47 | 0.68 | 0.18 | (12) | 1.304 | 3.66 | 1.19 | 0.19 |
| Eps Tau b | 594.90 | 13.70 | 3.66 | 189.32 | 1.46 | -0.04 | (32) | 1.010 | 12.64 | 3.61 | 0.11 |
| Eps. Eri b | 2,502.10 | | | 11.68 | 5.98 | -0.02 | (73) | 0.891 | 10.26 | 6.25 | -0.25 |
| Gam. Ceph.B | 903.30 | | | 68.02 | 2.37 | -0.13 | (73) | 1.030 | 10.82 | 4.37 | -0.13 |
| Gam. 1 Leo | 428.50 | 31.88 | 2.80 | 575.87 | 0.91 | -0.09 | (28) | 1.130 | 11.92 | 3.30 | -0.20 |
| GJ 1214 b | 1.58 | 0.21 | 2.00 | 5.31 | 0.67 | 0.17 | (73) | 1.730 | 0.68 | 1.32 | -0.18 |
| GJ 3021 b | 133.71 | 0.87 | 5.10 | 8.58 | 2.50 | 0.00 | (70) | 0.749 | 9.06 | 2.45 | -0.05 |
| GJ 436 b | 2.64 | 0.42 | 1.00 | 21.24 | 0.50 | 0.00 | (24) | 1.498 | 3.66 | 0.90 | -0.10 |
| GJ 581 b | 5.37 | 0.38 | 0.30 | 64.07 | 0.44 | -0.06 | (24) | 1.608 | 0.68 | 1.99 | -0.01 |
| GJ 674 b | 4.69 | 0.41 | 1.00 | 20.74 | 0.61 | 0.11 | (10) | 1.530 | 0.68 | 1.90 | -0.10 |
| GJ 676A b | 1,056.80 | 0.50 | 1.60 | 15.81 | 4.06 | 0.06 | (73) | 1.430 | 3.66 | 6.61 | 0.11 |
| GJ 785 b | 74.39 | 0.68 | 0.50 | 68.79 | 1.03 | 0.03 | (75) | 0.782 | 12.04 | 1.83 | -0.17 |
| GJ 849 b | 1,890.00 | 0.59 | 1.00 | 29.84 | 3.99 | -0.01 | (24) | 1.520 | 0.68 | 14.06 | 0.06 |
| Gl 581 c | 12.93 | 0.38 | 0.30 | 64.07 | 0.59 | 0.09 | (24) | 1.608 | 0.68 | 2.67 | 0.17 |
| Gl 581 d | 83.60 | 0.38 | 0.30 | 64.07 | 1.09 | 0.09 | (24) | 1.608 | 0.68 | 4.97 | -0.03 |
| Gl 581 e | 3.15 | 0.38 | 0.30 | 64.07 | 0.37 | -0.13 | (24) | 1.608 | 0.68 | 1.67 | 0.17 |
| Gl 649 b | 598.30 | 0.49 | 1.90 | 13.04 | 3.58 | 0.08 | (38) | 1.520 | 0.68 | 9.58 | 0.08 |
| Gl 86 b | 15.78 | 0.86 | 2.37 | 18.35 | 0.95 | -0.05 | (16) | 0.812 | 5.49 | 1.42 | -0.08 |
| Gl 876 b | 61.12 | 0.30 | 2.00 | 7.59 | 2.00 | 0.00 | (38) | 1.597 | 0.68 | 4.48 | -0.02 |
| Gl 876 c | 30.09 | 0.30 | 2.00 | 7.59 | 1.58 | 0.08 | (38) | 1.597 | 0.68 | 3.54 | 0.04 |
| Gl 876 d | 1.94 | 0.30 | 2.00 | 7.59 | 0.63 | 0.13 | (38) | 1.597 | 0.68 | 1.42 | -0.08 |
| Gl 876 e | 124.26 | 0.30 | 2.00 | 7.59 | 2.54 | 0.04 | (38) | 1.597 | 0.68 | 5.67 | 0.17 |
| Gl 179 b | 2,288 | 0.38 | 1.00 | 19.22 | 4.92 | -0.08 | (36) | 1.590 | 0.68 | 14.98 | -0.02 |
| HAT-P-11 b | 4.89 | 0.75 | 1.50 | 25.29 | 0.58 | 0.08 | (6) | 1.025 | 10.82 | 0.77 | -0.23 |
| HAT-P-12 b | 3.21 | 0.70 | 0.50 | 70.81 | 0.36 | -0.14 | (75) | | | | |
| HAT-P-13 b | 2.92 | 1.56 | 2.90 | 27.21 | 0.48 | -0.02 | (5) | 0.730 | 7.15 | 0.74 | 0.24 |
| HAT-P-13 c | 448.20 | 1.56 | 2.90 | 27.21 | 2.54 | 0.04 | (5) | 0.730 | 7.15 | 3.97 | -0.03 |
| HAT-P-14 b | 4.63 | 1.47 | 8.80 | 8.44 | 0.82 | -0.18 | (71) | 0.456 | 2.90 | 1.17 | 0.17 |
| HAT-P-15 b | 10.86 | 1.08 | 2.00 | 27.31 | 0.74 | 0.24 | (43) | 0.610 | 7.38 | 1.14 | 0.14 |
| HAT-P-16 b | 2.78 | 1.24 | 3.50 | 17.88 | 0.54 | 0.04 | (15) | 0.675 | 6.20 | 0.77 | -0.23 |
| HAT-P-17 b | 10.34 | 0.84 | 0.30 | 141.11 | 0.42 | -0.08 | (37) | 0.830 | 9.71 | 1.02 | 0.02 |
| HAT-P-17 c | 1,798 | 0.84 | 0.30 | 141.11 | 2.34 | -0.16 | (37) | 0.830 | 9.71 | 5.70 | 0.20 |
| Hat-P-18 b | 5.51 | 0.75 | 0.50 | 75.77 | 0.42 | -0.08 | (29) | 0.930 | 9.36 | 0.84 | -0.16 |
| Hat-P-19 b | 4.01 | 0.82 | 0.70 | 59.25 | 0.41 | -0.09 | (29) | 0.790 | 12.04 | 0.69 | 0.19 |
| HAT-P-1b | 4.47 | | | 15.34 | 0.66 | 0.16 | (73) | 0.575 | 4.72 | 0.98 | -0.02 |
| Hat-P-20 b | 2.88 | 0.69 | 2.10 | 16.72 | 0.56 | 0.06 | (6) | 1.231 | 14.14 | 0.59 | 0.09 |
| Hat-P-21 b | 4.12 | 1.11 | 3.50 | 15.97 | 0.64 | 0.14 | (6) | 1.093 | 11.92 | 0.70 | 0.20 |
| Hat-P-22 b | 3.12 | 1.04 | 0.50 | 105.20 | 0.31 | -0.19 | (6) | 0.826 | 9.71 | 0.69 | 0.19 |
| Hat-P-23 b | 1.21 | 1.20 | 8.10 | 7.51 | 0.54 | 0.04 | (6) | 1.106 | 11.92 | 0.47 | -0.03 |
| Hat-P-24 b | 3.36 | 1.32 | 11.50 | 5.79 | 0.83 | -0.17 | (41) | 0.450 | 2.90 | 1.05 | 0.05 |
| Hat-P-25 b | 3.65 | 0.96 | 0.50 | 97.01 | 0.34 | -0.16 | (75) | 0.609 | 5.10 | 0.89 | -0.11 |
| Hat-P-26 b | 4.23 | 0.79 | 1.80 | 22.14 | 0.58 | 0.08 | (75) | 0.530 | 5.87 | 0.90 | -0.10 |
| HAT-P-27 | 4.23 | 0.90 | 2.50 | 18.17 | 0.62 | 0.12 | (3) | | | | |
| HAT-P-2b | 5.63 | | | 4.05 | 1.12 | 0.12 | (73) | 0.423 | 1.67 | 1.50 | 0.00 |
| HAT-P-3b | 2.90 | 0.82 | 0.50 | 83.35 | 0.33 | -0.17 | (16) | 0.936 | 9.36 | 0.68 | 0.18 |
| HAT-P-4b | 3.06 | 1.59 | 5.50 | 14.62 | 0.59 | 0.09 | (42) | 0.771 | 5.61 | 0.82 | -0.18 |
| HAT-P-5b | 2.79 | 1.17 | 2.60 | 22.70 | 0.50 | 0.00 | (4) | 0.970 | 12.64 | 0.60 | 0.10 |
| HAT-P-6b | 3.85 | 1.46 | 8.70 | 8.49 | 0.77 | -0.23 | (16) | 0.670 | 6.20 | 0.85 | -0.15 |
| HAT-P-7b | 2.20 | 1.84 | 3.80 | 24.49 | 0.45 | -0.05 | (58) | 0.360 | 1.67 | 1.10 | 0.10 |
| HAT-P-8 b | 3.08 | 1.58 | 11.50 | 6.95 | 0.76 | -0.24 | (75) | 0.410 | 1.67 | 1.23 | 0.23 |
| HAT-P-9 b | 3.92 | 1.32 | 11.90 | 5.61 | 0.89 | -0.11 | (75) | 0.010 | 1.67 | 1.33 | -0.17 |
| HD 100777b | 383.70 | | | 39.00 | 2.14 | 0.14 | (73) | 0.760 | 9.06 | 3.49 | -0.01 |
| HD 10180 c | 5.76 | | | 24.00 | 0.62 | 0.12 | (44) | 0.629 | 6.45 | 0.96 | -0.04 |
| HD 10180 d | 16.36 | | | 24.00 | 0.88 | -0.12 | (44) | 0.629 | 6.45 | 1.36 | -0.14 |
| HD 10180 e | 49.75 | | | 24.00 | 1.28 | -0.22 | (44) | 0.629 | 6.45 | 1.98 | -0.02 |

| Exoplanet | Planet Orbital Period $P_n$ | Stellar Radius (Rs) | $v \sin i$ (km s⁻¹) | Present Rotation Period $P_{rot}$ | Present Orbital Ranks $n$ | $\Delta n$ | Ref. $v\sin i$ & $P_{rot}$ | B-V | Rotation Period $P_{rot}$ at 650 Myr | Orbital Ranks $n$ at 650 Myr | $\Delta n$ |
|---|---|---|---|---|---|---|---|---|---|---|---|
| HD 10180 f | 122.72 | | | 24.00 | 1.72 | 0.22 | (44) | 0.629 | 6.45 | 2.67 | 0.17 |
| HD 10180 g | 601.20 | | | 24.00 | 2.93 | -0.07 | (44) | 0.629 | 6.45 | 4.53 | 0.03 |
| HD 10180 h | 2,222 | | | 24.00 | 4.52 | 0.02 | (44) | 0.629 | 6.45 | 7.01 | 0.01 |
| HD 101930b | 70.46 | | | 46.00 | 1.15 | 0.15 | (73) | 0.909 | 10.26 | 1.90 | -0.10 |
| HD 102117b | 20.67 | | | 37.55 | 0.82 | -0.18 | (73) | 0.720 | 7.15 | 1.42 | -0.08 |
| HD 102195 b | 4,114 | | | 19.58 | 5.95 | -0.05 | (73) | 0.805 | 5.49 | 9.08 | -0.09 |
| HD 102272 b | 127.58 | 10.10 | 3.00 | 170.28 | 0.91 | -0.09 | (55) | 1.020 | 10.82 | 2.28 | -0.22 |
| HD 102272 c | 520.00 | 10.10 | 3.00 | 170.28 | 1.45 | -0.05 | (55) | 1.020 | 10.82 | 3.64 | 0.14 |
| HD 102365 b | 122.10 | 0.97 | 0.50 | 98.12 | 1.08 | 0.08 | (24) | 0.680 | 6.20 | 2.70 | 0.20 |
| HD 103197 b | 47.84 | | | 51.00 | 0.98 | -0.02 | (50) | 0.860 | 9.18 | 1.73 | 0.23 |
| HD 104985b | 198.20 | | | 120.98 | 1.18 | 0.18 | (73) | 1.030 | 10.82 | 2.64 | 0.14 |
| HD 106252b | 1,516 | | | 22.80 | 4.05 | 0.05 | (73) | 0.635 | 8.65 | 5.60 | 0.08 |
| HD 10647b | 1,003 | | | 7.90 | 5.03 | 0.03 | (73) | 0.550 | 3.00 | 6.94 | 0.02 |
| HD 10697b | 1,072.30 | | | 36.00 | 3.10 | 0.10 | (73) | 0.720 | 5.02 | 5.98 | -0.02 |
| HD 107148b | 48.06 | | | 33.00 | 1.13 | 0.13 | (73) | 0.707 | 7.25 | 1.88 | -0.12 |
| HD 108147b | 10.90 | | | 8.70 | 1.08 | 0.08 | (73) | 0.553 | 3.00 | 1.54 | 0.04 |
| HD 108874 b | 395.40 | | | 40.40 | 2.14 | 0.14 | (73) | 0.738 | 9.13 | 3.51 | 0.01 |
| HD 108874 c | 1,605.80 | | | 40.40 | 3.41 | -0.09 | (73) | 0.738 | 9.13 | 5.60 | 0.10 |
| HD 109246b | 68.27 | 1.02 | 3.00 | 17.20 | 1.58 | 0.08 | (9) | 0.640 | 8.65 | 1.99 | -0.01 |
| HD 109749 b | 5.24 | | | 34.00 | 0.54 | 0.04 | (73) | 0.680 | 6.20 | 0.95 | -0.05 |
| HD 111232b | 1,143.00 | | | 30.70 | 3.34 | -0.16 | (73) | 0.701 | 7.25 | 5.40 | -0.10 |
| HD 114386b | 937.70 | | | 35.57 | 2.98 | -0.02 | (73) | 1.042 | 10.82 | 4.43 | -0.07 |
| HD 114729b | 1,131.48 | | | 18.57 | 3.94 | -0.06 | (73) | 0.591 | 5.64 | 5.85 | -0.15 |
| HD 114762b | 83.89 | 1.12 | 1.77 | 32.06 | 1.38 | -0.12 | (16) | 0.529 | 5.87 | 2.43 | -0.07 |
| HD 114783b | 501.00 | | | 45.20 | 2.23 | 0.23 | (73) | 0.902 | 10.26 | 3.66 | 0.16 |
| HD 11506 c | 170.46 | | | 18.30 | 2.10 | 0.10 | (73) | 0.680 | 6.20 | 3.02 | 0.02 |
| HD 11506b | 1,280 | | | 18.30 | 4.12 | 0.12 | (73) | 0.680 | 6.20 | 5.91 | -0.09 |
| HD 117207b | 2,627.08 | | | 36.00 | 4.18 | 0.18 | (73) | 0.716 | 7.25 | 7.13 | 0.13 |
| HD 117618b | 52.20 | 1.20 | 3.19 | 19.03 | 1.40 | -0.10 | (16) | 0.598 | 5.64 | 2.10 | 0.10 |
| HD 118203b | 6.13 | 1.96 | 4.70 | 21.09 | 0.66 | 0.16 | (16) | 0.682 | 6.20 | 1.00 | 0.00 |
| HD 11964 b | 37.82 | | | 49.00 | 0.92 | -0.08 | (73) | 0.817 | 5.49 | 1.90 | -0.10 |
| HD 11964 c | 2,110 | | | 49.00 | 3.51 | 0.01 | (73) | 0.817 | 5.49 | 7.27 | 0.07 |
| HD 11977b | 711.00 | 10.47 | 2.40 | 220.68 | 1.48 | -0.02 | (16) | 0.913 | 10.26 | 4.11 | 0.11 |
| HD 121504b | 63.33 | | | 8.60 | 1.95 | -0.05 | (73) | 0.593 | 5.64 | 2.24 | 0.24 |
| HD 122430b | 344.95 | 22.90 | 4.70 | 246.44 | 1.12 | 0.12 | (64) | 1.331 | 3.66 | 4.55 | 0.05 |
| HD 125612 b | 502.00 | | | 17.63 | 3.05 | 0.05 | (73) | 0.574 | 4.72 | 4.74 | -0.23 |
| HD 125612 c | 4.15 | | | 17.63 | 0.62 | 0.12 | (73) | 0.574 | 4.72 | 0.96 | -0.04 |
| HD 125612 d | 4,613 | | | 17.63 | 6.40 | -0.10 | (73) | 0.574 | 4.72 | 9.92 | -0.08 |
| HD 12661 b | 263.60 | | | 35.00 | 1.96 | -0.04 | (73) | 0.710 | 7.15 | 3.33 | -0.17 |
| HD 12661 c | 1,444.50 | | | 35.00 | 3.46 | -0.04 | (73) | 0.710 | 7.15 | 5.87 | -0.13 |
| HD 126614 b | 1,244 | 1.09 | 2.00 | 27.57 | 3.56 | 0.06 | (75) | 1.200 | 14.14 | 4.45 | -0.05 |
| HD 128311 b | 448.60 | | | 10.78 | 3.47 | -0.03 | (73) | 1.026 | 10.82 | 3.46 | -0.04 |
| HD 128311 c | 919.00 | | | 10.78 | 4.40 | -0.10 | (73) | 1.026 | 10.82 | 4.40 | -0.10 |
| HD 130322 b | 10.72 | | | 8.70 | 1.07 | 0.07 | (73) | 0.778 | 5.61 | 1.24 | 0.24 |
| HD 131664 b | 1,951 | | | 22.00 | 4.46 | -0.04 | (51) | 0.667 | 8.64 | 6.09 | 0.09 |
| HD 13189b | 471.60 | 3.62 | 2.40 | 76.29 | 1.84 | -0.16 | (63) | 1.465 | 3.66 | 5.05 | 0.05 |
| HD 132406 b | 974.00 | 1.23 | 1.70 | 36.60 | 2.99 | -0.01 | (19) | 0.650 | 7.73 | 5.01 | 0.01 |
| HD 134987 b | 258.19 | | | 30.50 | 2.04 | 0.04 | (73) | 0.672 | 6.20 | 3.47 | -0.03 |
| HD 134987 c | 5,000 | | | 30.50 | 5.47 | -0.03 | (73) | 0.672 | 6.20 | 9.31 | -0.19 |
| HD 136118b | 1,209.60 | | | 13.00 | 4.53 | 0.03 | (73) | 0.521 | 5.87 | 5.91 | -0.09 |
| HD 136418 b | 464.30 | 3.40 | 1.66 | 103.59 | 1.65 | 0.15 | (75) | 0.930 | 9.36 | 3.67 | 0.17 |
| HD 137510 b | 804.90 | 2.01 | 7.98 | 12.77 | 3.98 | -0.02 | (75) | 0.618 | 7.38 | 4.78 | -0.22 |
| HD 137759b | 511.10 | 1.60 | 1.50 | 53.95 | 2.12 | 0.12 | (24) | 1.170 | 11.92 | 3.50 | 0.00 |
| HD 13931 b | 4,218 | 1.23 | 2.02 | 30.80 | 5.15 | 0.15 | (75) | 0.640 | 8.65 | 7.87 | -0.13 |
| HD 141937b | 653.22 | | | 21.00 | 3.14 | 0.14 | (73) | 0.584 | 5.64 | 4.87 | -0.13 |
| HD 142 b | 337.11 | | | 10.75 | 3.15 | 0.15 | (73) | 0.519 | 2.50 | 5.13 | 0.13 |

| Exoplanet | Planet Orbital Period $P_n$ | Stellar Radius (Rs) | $v \sin i$ (km s⁻¹) | Present Rotation Period $P_{rot}$ | Present Orbital Ranks $n$ | $\Delta n$ | Ref. vsini & $P_{rot}$ | B-V | Rotation Period $P_{rot}$ at 650 Myr | Orbital Ranks $n$ at 650 Myr | $\Delta n$ |
|---|---|---|---|---|---|---|---|---|---|---|---|
| HD 142022A | 1,928 | 1.02 | 1.20 | 43.13 | 3.55 | 0.05 | (20) | 0.775 | 5.61 | 7.00 | 0.00 |
| HD 142415b | 386.30 | | | 13.87 | 3.03 | 0.03 | (46) | 0.621 | 8.65 | 3.55 | 0.05 |
| HD 145377 b | 103.95 | | | 12.00 | 2.05 | 0.05 | (51) | 0.631 | 6.45 | 2.53 | 0.03 |
| HD 1461 b | 5.77 | 1.10 | 1.60 | 34.61 | 0.55 | 0.05 | (75) | 0.674 | 6.20 | 0.98 | -0.02 |
| HD 147018 b | 44.24 | 0.99 | 1.56 | 32.03 | 1.11 | 0.11 | (75) | 0.763 | 5.61 | 1.99 | -0.01 |
| HD 147018 c | 1,008 | 0.99 | 1.56 | 32.03 | 3.16 | 0.16 | (75) | 0.763 | 5.61 | 5.64 | 0.14 |
| HD 147513b | 528.40 | | | 8.53 | 3.96 | -0.04 | (73) | 0.625 | 8.65 | 3.94 | -0.06 |
| HD 148156 b | 1,010 | 1.21 | 5.70 | 10.74 | 4.55 | 0.05 | (53) | 0.558 | 3.00 | 6.96 | -0.04 |
| HD 148427 b | 331.50 | 3.22 | 2.10 | 77.55 | 1.62 | 0.12 | (75) | 0.930 | 9.36 | 3.28 | -0.22 |
| HD 149026b | 2.88 | 1.50 | 6.00 | 12.62 | 0.61 | 0.11 | (16) | 0.600 | 5.10 | 0.83 | -0.17 |
| HD 149143b | 4.07 | | | 28.00 | 0.53 | 0.03 | (73) | 0.673 | 6.20 | 0.87 | -0.13 |
| HD 150706b | 264.00 | | | 9.37 | 3.04 | 0.04 | (73) | 0.614 | 7.38 | 3.29 | -0.21 |
| HD 153950 b | 499.40 | | | 14.00 | 3.29 | -0.21 | (51) | 0.565 | 4.00 | 5.00 | 0.00 |
| HD 154345b | 10,900.00 | 0.89 | 1.50 | 30.11 | 7.13 | 0.13 | (25) | 0.730 | 7.15 | 11.51 | 0.01 |
| HD 154672 b | 163.94 | 1.27 | 0.54 | 118.95 | 1.11 | 0.11 | (75) | 0.710 | 7.25 | 2.83 | -0.17 |
| HD 154857b | 409.00 | | | 30.90 | 2.37 | -0.13 | (73) | 0.684 | 6.20 | 4.04 | 0.04 |
| HD 155358 b | 195.00 | 1.15 | 2.00 | 29.04 | 1.89 | -0.11 | (24) | 0.549 | 3.00 | 4.02 | 0.02 |
| HD 155358 c | 530.30 | 1.15 | 2.00 | 29.04 | 2.63 | 0.13 | (56) | 0.549 | 3.00 | 5.61 | 0.11 |
| HD 156411 b | 842.20 | 2.16 | 3.30 | 33.11 | 2.94 | -0.06 | (53) | 0.614 | 7.38 | 4.85 | -0.15 |
| HD 156668 b | 4.65 | 0.72 | 0.50 | 72.83 | 0.40 | -0.10 | (36) | 1.010 | 12.64 | 0.72 | 0.22 |
| HD 156846 b | 359.51 | 1.91 | 4.45 | 21.66 | 2.55 | 0.05 | (75) | 0.568 | 4.00 | 4.48 | -0.02 |
| HD 159868b | 986.00 | | | 37.62 | 2.97 | -0.03 | (73) | 0.698 | 7.25 | 5.14 | 0.14 |
| HD 160691 b | 643.25 | | | 31.81 | 2.72 | 0.22 | (73) | 0.694 | 7.25 | 4.46 | -0.04 |
| HD 160691 c | 9.64 | | | 31.81 | 0.67 | 0.17 | (73) | 0.694 | 7.25 | 1.10 | 0.10 |
| HD 160691 d | 310.55 | | | 31.81 | 2.14 | 0.14 | (73) | 0.694 | 7.25 | 3.50 | 0.00 |
| HD 160691 e | 4,205.80 | | | 31.81 | 5.09 | 0.09 | (73) | 0.694 | 7.25 | 8.34 | 7.34 |
| HD 16141b | 75.82 | 1.35 | 3.00 | 22.74 | 1.49 | -0.01 | (56) | 0.669 | 8.64 | 2.06 | 0.06 |
| HD 16175 b | 990.00 | 1.87 | 5.50 | 17.20 | 3.86 | -0.14 | (25) | 0.630 | 8.65 | 4.86 | -0.14 |
| HD 162020b | 8.43 | | | 1.62 | 1.73 | 0.23 | (73) | 0.964 | 9.36 | 0.97 | -0.03 |
| HD 16417 b | 17.24 | 1.63 | 1.20 | 68.70 | 0.63 | 0.13 | (75) | 0.670 | 8.64 | 1.26 | -0.24 |
| HD 164922 b | 1,155.00 | | | 43.00 | 2.99 | -0.01 | (73) | 0.799 | 12.04 | 4.58 | 0.08 |
| HD 167042 b | 416.10 | 4.30 | 2.50 | 87.00 | 1.68 | 0.18 | (16) | 0.940 | 9.36 | 3.54 | 0.04 |
| HD 16760 b | 465.10 | 0.91 | 2.80 | 16.47 | 3.04 | 0.04 | (75) | 0.710 | 7.25 | 4.00 | 0.00 |
| HD 167665 b | 4,385 | 1.10 | 5.20 | 10.70 | 7.43 | -0.07 | (59) | 0.536 | 5.13 | 9.49 | -0.01 |
| HD 168443 b | 58.12 | | | 38.61 | 1.15 | 0.15 | (73) | 0.741 | 9.06 | 1.86 | -0.14 |
| HD 168443 c | 1,739.50 | | | 38.61 | 3.56 | 0.06 | (73) | 0.724 | 5.02 | 7.02 | 0.02 |
| HD 168746 b | 6.43 | | | 34.77 | 0.57 | 0.07 | (73) | 0.710 | 7.25 | 0.96 | -0.04 |
| HD 1690 b | 533.00 | 16.70 | 3.50 | 241.33 | 1.30 | -0.20 | (52) | 1.354 | 3.66 | 5.26 | -0.24 |
| HD 169822 b | 293.10 | 0.95 | 6.60 | 7.28 | 3.43 | -0.07 | (24) | 0.699 | 7.25 | 3.43 | -0.07 |
| HD 169830 b | 225.62 | | | 8.30 | 3.01 | 0.01 | (73) | 0.518 | 2.50 | 4.49 | -0.01 |
| HD 169830 c | 2,102 | | | 8.30 | 6.33 | -0.17 | (73) | 0.518 | 2.50 | 9.44 | -0.06 |
| HD 170469 b | 1,145 | | | 13.00 | 4.45 | -0.05 | (73) | 0.990 | 12.64 | 4.49 | -0.01 |
| HD 17051 b | 311.30 | | | 7.27 | 3.50 | 0.00 | (73) | 0.570 | 4.72 | 4.04 | 0.04 |
| HD 17092 b | 359.90 | | | 505.00 | 0.89 | -0.11 | (54) | 1.260 | 14.14 | 2.94 | -0.06 |
| HD 171028 | 538.00 | 1.95 | 2.30 | 42.88 | 2.32 | -0.18 | (16) | 0.644 | 8.65 | 3.96 | -0.04 |
| HD 171238 b | 1,523.00 | 0.91 | 1.48 | 31.20 | 3.65 | 0.15 | (73) | 0.740 | 9.13 | 5.50 | 0.00 |
| HD 17156 b | 21.22 | 1.47 | 2.60 | 28.60 | 0.91 | -0.09 | (16) | 0.627 | 6.45 | 1.49 | -0.01 |
| HD 175541 b | 297.30 | | | 59.98 | 1.71 | 0.21 | (73) | 0.885 | 10.26 | 3.07 | 0.07 |
| HD 177830 b | 391.60 | | | 65.00 | 1.82 | -0.18 | (73) | 1.062 | 10.82 | 3.31 | -0.19 |
| HD 177830 c | 110.90 | | | 65.00 | 1.19 | 0.19 | (73) | 1.062 | 10.82 | 2.17 | 0.17 |
| HD178911B b | 71.49 | | | 36.00 | 1.26 | -0.24 | (73) | 0.750 | 9.06 | 1.99 | -0.01 |
| HD 179079 b | 14.48 | 1.48 | 0.50 | 149.71 | 0.46 | -0.04 | (75) | 0.744 | 9.06 | 1.17 | 0.17 |
| HD 179949 b | 3.09 | | | 10.00 | 0.68 | 0.18 | (73) | 0.548 | 3.00 | 1.01 | 0.01 |
| HD 180902 b | 479.00 | 4.10 | 2.88 | 72.00 | 1.88 | -0.12 | (40) | 0.940 | 9.36 | 3.71 | 0.21 |
| HD 181342 b | 663.00 | 4.60 | 3.04 | 76.53 | 2.05 | 0.05 | (40) | 1.020 | 10.82 | 3.94 | -0.06 |
| HD 181433 b | 9.37 | | | 54.00 | 0.56 | 0.06 | (13) | 1.010 | 12.64 | 0.91 | -0.09 |

| Exoplanet | Planet Orbital Period $P_n$ | Stellar Radius (Rs) | $v \sin i$ (km s⁻¹) | Present Rotation Period $P_{rot}$ | Present Orbital Ranks $n$ | $\Delta n$ | Ref. vsini & $P_{rot}$ | B-V | Rotation Period $P_{rot}$ at 650 Myr | Orbital Ranks $n$ at 650 Myr | $\Delta n$ |
|---|---|---|---|---|---|---|---|---|---|---|---|
| HD 181433 c | 962.00 | | | 54.00 | 2.61 | 0.11 | (13) | 1.010 | 12.64 | 4.24 | 0.24 |
| HD 181433 d | 2,172 | | | 54.00 | 3.43 | -0.07 | (13) | 1.010 | 12.64 | 5.56 | 0.06 |
| HD 181720 b | 956.00 | 1.46 | 2.00 | 36.85 | 2.96 | -0.04 | (24) | 0.599 | 5.64 | 5.53 | 0.03 |
| HD 183263 b | 634.23 | | | 32.00 | 2.71 | 0.21 | (73) | 0.678 | 6.20 | 4.68 | 0.18 |
| HD 183263 c | 3,066 | | | 32.00 | 4.58 | 0.08 | (73) | 0.678 | 6.20 | 7.91 | -0.09 |
| HD 185269 b | 6.84 | | | 23.00 | 0.67 | 0.17 | (73) | 0.606 | 5.10 | 1.10 | 0.10 |
| HD 187085 b | 986.00 | | | 14.35 | 4.10 | 0.10 | (73) | 0.574 | 4.72 | 5.93 | -0.07 |
| HD 187123 b | 3.10 | | | 30.00 | 0.47 | -0.03 | (73) | 0.645 | 8.65 | 0.71 | 0.21 |
| HD 187123 c | 3,700 | | | 30.00 | 4.98 | -0.02 | (73) | 0.645 | 8.65 | 7.53 | 0.03 |
| HD 188015 b | 456.46 | 1.10 | 2.00 | 27.82 | 2.54 | 0.04 | (56) | 0.730 | 7.15 | 4.00 | 0.00 |
| HD 189733 b | 2.22 | | | 19.75 | 0.48 | -0.02 | (73) | 0.932 | 14.40 | 0.54 | 0.04 |
| HD 190228 b | 1,146.00 | | | 47.00 | 2.90 | -0.10 | (73) | 0.793 | 12.04 | 4.57 | 0.07 |
| HD 190360 b | 2,891.00 | | | 40.00 | 4.17 | 0.17 | (73) | 0.761 | 5.61 | 8.02 | 0.02 |
| HD 190360 c | 17.10 | | | 35.87 | 0.78 | -0.22 | (73) | 0.761 | 5.61 | 1.45 | -0.05 |
| HD 190647 b | 1,038.10 | | | 39.00 | 2.99 | -0.01 | (73) | 0.743 | 9.06 | 4.86 | -0.14 |
| HD 190984 b | 4,885 | 1.53 | 3.40 | 22.76 | 5.99 | -0.01 | (75) | 0.579 | 4.72 | 10.12 | 0.12 |
| HD 192263 b | 24.36 | | | 23.98 | 1.01 | 0.01 | (73) | 0.940 | 9.36 | 1.38 | -0.12 |
| HD 192699 b | 351.50 | | | 59.75 | 1.81 | -0.19 | (73) | 0.858 | 9.18 | 3.37 | -0.13 |
| HD 195019 b | 18.27 | | | 22.00 | 0.94 | -0.06 | (73) | 0.674 | 6.20 | 1.43 | -0.07 |
| HD 196050 b | 1,300 | | | 28.60 | 3.57 | 0.07 | (73) | 0.656 | 7.43 | 5.59 | 0.09 |
| HD 196885 b | 386.00 | | | 15.00 | 2.95 | -0.05 | (73) | 0.560 | 4.00 | 4.59 | 0.09 |
| HD 19994 b | 454.00 | | | 10.78 | 3.48 | -0.02 | (73) | 0.575 | 4.72 | 4.58 | 0.08 |
| HD 200964 b | 630.00 | 4.30 | 2.28 | 95.39 | 1.88 | -0.12 | (40) | 0.880 | 10.26 | 3.95 | -0.05 |
| HD 200964 c | 825.00 | 4.30 | 2.28 | 95.39 | 2.05 | 0.05 | (40) | 0.880 | 10.26 | 4.32 | -0.18 |
| HD 202206 b | 255.87 | | | 22.98 | 2.23 | 0.23 | (73) | 0.721 | 6.20 | 3.46 | -0.04 |
| HD 202206 c | 1,296.80 | | | 22.98 | 3.84 | -0.16 | (73) | 0.721 | 6.20 | 5.94 | -0.06 |
| HD 20367 b | 500.00 | | | 5.47 | 4.50 | 0.00 | (73) | 0.563 | 4.00 | 5.00 | 0.00 |
| HD 2039 b | 1,192.58 | | | 26.20 | 3.57 | 0.07 | (73) | 0.656 | 7.43 | 5.43 | -0.07 |
| HD 204313 b | 1,931 | 1.10 | 1.80 | 30.82 | 3.97 | -0.03 | (24) | 0.697 | 7.25 | 6.43 | -0.07 |
| HD 205739 b | 279.80 | 1.33 | 4.48 | 15.02 | 2.65 | 0.15 | (75) | 0.550 | 3.00 | 4.53 | 0.03 |
| HD 206610 b | 610.00 | 6.10 | 3.30 | 93.49 | 1.87 | -0.13 | (73) | 1.010 | 12.64 | 3.64 | 0.14 |
| HD 20782b | 585.86 | | | 21.18 | 3.02 | 0.02 | (73) | 0.627 | 6.45 | 4.50 | 0.00 |
| HD 208487b | 130.00 | | | 12.51 | 2.18 | 0.18 | (73) | 0.570 | 4.72 | 3.02 | 0.02 |
| HD 20868 b | 380.85 | | | 51.00 | 1.95 | -0.05 | (51) | 1.037 | 10.82 | 3.28 | -0.22 |
| HD 209458b | 3.52 | | | 19.00 | 0.57 | 0.07 | (73) | 0.574 | 4.72 | 0.91 | -0.09 |
| HD 210277b | 435.60 | | | 40.80 | 2.20 | 0.20 | (73) | 0.759 | 9.06 | 3.64 | 0.14 |
| HD 210702b | 341.10 | | | 69.09 | 1.70 | 0.20 | (73) | 0.955 | 12.64 | 3.00 | 0.00 |
| HD 212301b | 2.25 | | | 12.00 | 0.57 | 0.07 | (73) | 0.551 | 3.00 | 0.91 | -0.09 |
| HD 212771 b | 373.30 | 5.00 | 2.63 | 96.16 | 1.57 | 0.07 | (40) | 0.880 | 10.26 | 3.31 | -0.19 |
| HD 213240b | 951.00 | | | 15.00 | 3.99 | -0.01 | (73) | 0.612 | 7.38 | 5.05 | 0.05 |
| HD 215497 b | 3.93 | 1.01 | 1.67 | 30.59 | 0.50 | 0.00 | (75) | 0.953 | 9.36 | 0.75 | 0.25 |
| HD 215497 c | 567.94 | 1.01 | 1.67 | 30.59 | 2.65 | 0.15 | (75) | 0.953 | 9.36 | 3.93 | -0.07 |
| HD 216435 b | 1,391 | | | 21.60 | 4.01 | 0.01 | (73) | 0.621 | 6.45 | 6.00 | 0.00 |
| HD 216437 b | 1,256 | | | 27.23 | 3.59 | 0.09 | (73) | 0.660 | 7.43 | 5.53 | 0.03 |
| HD 216770 b | 118.45 | | | 35.60 | 1.49 | -0.01 | (46) | 0.850 | 9.18 | 2.35 | -0.15 |
| HD 217107 b | 7.13 | 1.12 | 1.70 | 33.38 | 0.60 | 0.10 | (24) | 0.744 | 9.06 | 0.92 | -0.08 |
| HD 217107 c | 3,150 | 1.12 | 1.70 | 33.38 | 4.55 | 0.05 | (24) | 0.744 | 9.06 | 7.03 | 0.03 |
| HD 217786 b | 1,319 | 1.27 | 1.40 | 45.88 | 3.06 | 0.06 | (52) | 0.578 | 4.72 | 6.54 | 0.04 |
| HD 219449b | 181.84 | 6.70 | 5.70 | 59.45 | 1.45 | -0.05 | (69) | 1.107 | 11.92 | 2.48 | -0.02 |
| HD 219828 b | 3.83 | | | 26.00 | 0.53 | 0.03 | (73) | 0.654 | 7.43 | 0.80 | -0.20 |
| HD 221287b | 456.10 | | | 5.00 | 4.50 | 0.00 | (73) | 0.502 | 3.70 | 4.98 | -0.02 |
| HD 222582b | 575.90 | | | 25.00 | 2.85 | -0.15 | (73) | 0.648 | 8.65 | 4.05 | 0.05 |
| HD 224693b | 26.73 | | | 27.40 | 0.99 | -0.01 | (73) | 0.640 | 8.65 | 1.46 | -0.04 |
| HD 23079b | 738.46 | | | 17.46 | 3.48 | -0.02 | (73) | 0.583 | 5.64 | 5.08 | 0.08 |
| HD 23127 b | 1,214 | | | 32.03 | 3.36 | -0.14 | (73) | 0.690 | 7.25 | 5.51 | 0.01 |
| HD 231701b | 141.60 | | | 10.28 | 2.40 | -0.10 | (73) | 0.539 | 5.13 | 3.02 | 0.02 |

| Exoplanet | Planet Orbital Period $P_n$ | Stellar Radius (Rs) | $v \sin i$ (km s⁻¹) | Present Rotation Period $P_{rot}$ | Present Orbital Ranks $n$ | $\Delta n$ | Ref. $v\sin i$ & $P_{rot}$ | B-V | Rotation Period $P_{rot}$ at 650 Myr | Orbital Ranks $n$ at 650 Myr | $\Delta n$ |
|---|---|---|---|---|---|---|---|---|---|---|---|
| HD 23596b | 1,565 | | | 25.00 | 3.97 | -0.03 | (73) | 0.618 | 7.38 | 5.96 | -0.04 |
| HD 240210 b | 501.75 | | | 654.00 | 0.92 | -0.08 | (76) | 1.630 | 0.68 | 9.04 | 0.04 |
| HD 25171 b | 1,845 | 1.18 | 1.00 | 59.68 | 3.14 | 0.14 | (52) | 0.554 | 3.00 | 8.50 | 0.00 |
| HD 2638b | 3.44 | | | 37.00 | 0.45 | -0.05 | (73) | 1.004 | 12.64 | 0.65 | 0.15 |
| HD 27442b | 423.84 | | | 119.85 | 1.52 | 0.02 | (73) | 1.073 | 10.82 | 3.40 | -0.10 |
| HD 27894 b | 17.99 | | | 44.00 | 0.74 | 0.24 | (73) | 1.003 | 12.64 | 1.12 | 0.12 |
| HD 28185b | 383.00 | | | 27.18 | 2.42 | -0.08 | (73) | 0.734 | 9.13 | 3.47 | -0.03 |
| HD 28254 b | 1,116 | 1.48 | 2.50 | 29.94 | 3.34 | -0.16 | (53) | 0.722 | 7.15 | 5.38 | -0.12 |
| HD 290327 b | 2,443 | 1.00 | 1.44 | 35.12 | 4.11 | 0.11 | (53) | 0.761 | 5.61 | 7.58 | 0.08 |
| HD 30177b | 2,819.65 | | | 44.41 | 3.99 | -0.01 | (73) | 0.773 | 5.61 | 7.95 | -0.05 |
| HD 30562 b | 1,157 | 1.64 | 4.32 | 19.17 | 3.92 | -0.08 | (75) | 0.629 | 8.65 | 5.11 | 0.11 |
| HD 31253 b | 466.00 | 1.71 | 3.00 | 28.83 | 2.53 | 0.03 | (24) | 0.578 | 4.72 | 4.62 | 0.12 |
| HD 32518 b | 157.54 | 10.22 | 1.10 | 469.92 | 0.69 | 0.19 | (24) | 1.107 | 11.92 | 2.36 | -0.14 |
| HD 330075 b | 3.39 | | | 48.00 | 0.41 | -0.09 | (73) | 0.932 | 19.19 | 0.56 | 0.06 |
| HD 33283b | 18.18 | | | 55.50 | 0.69 | 0.19 | (73) | 0.586 | 5.64 | 1.48 | -0.02 |
| HD 33564b | 388.00 | | | 8.86 | 3.52 | 0.02 | (73) | 0.513 | 2.50 | 5.37 | -0.13 |
| HD 33636b | 2,828 | | | 16.17 | 5.59 | 0.09 | (73) | 0.588 | 5.64 | 7.94 | -0.06 |
| HD 34445 b | 1,049 | 1.38 | 2.70 | 25.85 | 3.44 | -0.06 | (36) | 0.659 | 7.43 | 5.21 | 0.21 |
| HD 3651b | 62.23 | | | 44.00 | 1.12 | 0.12 | (73) | 0.850 | 9.18 | 1.89 | -0.11 |
| HD 37124 b | 154.46 | | | 25.00 | 1.83 | -0.17 | (73) | 0.667 | 8.64 | 2.61 | 0.11 |
| HD 37124 c | 885.50 | | | 27.31 | 3.19 | 0.19 | (73) | 0.667 | 8.64 | 4.68 | 0.18 |
| HD 37124 d | 2,295 | | | 25.00 | 4.51 | 0.01 | (73) | 0.667 | 8.64 | 6.43 | -0.07 |
| HD 37605b | 55.23 | 0.87 | 3.50 | 12.59 | 1.64 | 0.14 | (25) | 0.834 | 9.18 | 1.82 | -0.18 |
| HD 38529 b | 14.31 | | | 35.00 | 0.74 | 0.24 | (73) | 0.773 | 5.61 | 1.37 | -0.13 |
| HD 38529 c | 2,134.76 | | | 35.00 | 3.94 | -0.06 | (73) | 0.773 | 5.61 | 7.25 | -0.21 |
| HD 38801 b | 696.30 | 2.53 | 0.54 | 236.97 | 1.43 | -0.07 | (75) | 0.873 | 19.19 | 3.31 | -0.19 |
| HD 39091b | 2,063.82 | | | 17.33 | 4.92 | -0.08 | (73) | 0.600 | 5.10 | 7.40 | -0.10 |
| HD 40307 b | 4.31 | | | 48.00 | 0.45 | -0.05 | (47) | 0.930 | 9.36 | 0.77 | -0.23 |
| HD 40307 c | 9.62 | | | 48.00 | 0.59 | 0.09 | (47) | 0.930 | 9.36 | 1.01 | 0.01 |
| HD 40307 d | 20.46 | | | 48.00 | 0.75 | -0.25 | (47) | 0.930 | 9.36 | 1.30 | -0.20 |
| HD 40979 b | 267.20 | | | 9.00 | 3.10 | 0.10 | (73) | 0.562 | 4.00 | 4.06 | 0.06 |
| HD 41004A b | 963.00 | 0.91 | 1.22 | 37.81 | 2.94 | -0.06 | (62) | 0.890 | 10.26 | 4.54 | -0.02 |
| HD 41004B b | 1.33 | 0.91 | 2.00 | 23.06 | 0.39 | -0.11 | (24) | 0.887 | 10.26 | 0.51 | 0.01 |
| HD 4113 b | 526.62 | 1.07 | 1.37 | 39.50 | 2.37 | -0.13 | (73) | 0.716 | 7.15 | 4.19 | 0.19 |
| HD 4203b | 404.22 | | | 45.00 | 2.08 | 0.08 | (73) | 0.757 | 9.06 | 3.55 | 0.05 |
| HD 4208b | 829.00 | 0.85 | 3.00 | 14.33 | 3.87 | -0.13 | (24) | 0.664 | 8.64 | 4.58 | 0.08 |
| HD 4308b | 15.56 | | | 25.15 | 0.85 | -0.15 | (73) | 0.655 | 7.43 | 1.28 | -0.22 |
| HD 4313 b | 356.00 | 4.90 | 2.76 | 89.80 | 1.58 | 0.08 | (40) | 0.960 | 9.36 | 3.36 | -0.14 |
| HD 43197 b | 327.80 | 1.00 | 2.18 | 23.20 | 2.42 | -0.08 | (53) | 0.817 | 5.49 | 3.91 | -0.09 |
| HD 43691b | 36.96 | 1.90 | 6.00 | 16.01 | 1.32 | -0.18 | (53) | 0.650 | 7.73 | 1.68 | 0.18 |
| HD 44219 b | 472.30 | 1.32 | 2.22 | 30.07 | 2.50 | 0.00 | (53) | 0.690 | 7.25 | 4.02 | 0.02 |
| HD 45350b | 890.76 | | | 39.00 | 2.84 | -0.16 | (73) | 0.740 | 9.13 | 4.60 | 0.10 |
| HD 45364 b | 226.93 | 0.85 | 1.80 | 23.88 | 2.12 | 0.12 | (24) | 0.720 | 7.15 | 3.17 | 0.17 |
| HD 45364 c | 342.85 | 0.85 | 1.00 | 42.99 | 2.00 | 0.00 | (75) | 0.720 | 7.15 | 3.63 | 0.13 |
| HD 45652 b | 43.60 | 0.93 | 8.50 | 5.55 | 1.99 | -0.01 | (24) | 0.850 | 9.18 | 1.68 | 0.18 |
| HD 46375 b | 3.02 | | | 43.00 | 0.41 | -0.09 | (73) | 0.871 | 19.19 | 0.54 | 0.04 |
| HD 47186 b | 4.08 | | | 33.00 | 0.50 | 0.00 | (13) | 0.710 | 7.25 | 0.83 | -0.17 |
| HD 47186 c | 1,353.60 | | | 33.00 | 3.45 | -0.05 | (13) | 0.710 | 7.25 | 5.72 | 0.22 |
| HD 47536 b | 712.13 | 23.47 | 1.90 | 624.78 | 1.04 | 0.04 | (16) | 1.180 | 11.92 | 3.91 | -0.09 |
| HD 47536 c | 2,500 | 23.47 | 1.90 | 624.78 | 1.59 | 0.09 | (16) | 1.180 | 11.92 | 5.94 | -0.06 |
| HD 49674 b | 4.95 | | | 27.30 | 0.57 | 0.07 | (73) | 0.729 | 5.02 | 1.00 | 0.00 |
| HD 50499 b | 2,582.70 | | | 21.00 | 4.97 | -0.03 | (73) | 0.610 | 7.38 | 7.05 | 0.05 |
| HD 50554 b | 1,293.00 | | | 14.67 | 4.45 | -0.05 | (73) | 0.571 | 4.72 | 6.49 | -0.01 |
| HD 52265 b | 119.60 | | | 14.60 | 2.02 | 0.02 | (73) | 0.572 | 4.72 | 2.94 | -0.06 |
| HD 5319 b | 675.00 | 3.26 | 3.31 | 49.81 | 2.38 | -0.12 | (75) | 0.607 | 5.10 | 5.10 | 0.10 |
| HD 5388 b | 777.00 | 1.58 | 4.20 | 19.09 | 3.44 | -0.06 | (75) | 0.500 | 3.70 | 5.94 | -0.06 |

| Exoplanet | Planet Orbital Period $P_n$ | Stellar Radius (Rs) | $v \sin i$ (km s$^{-1}$) | Present Rotation Period $P_{rot}$ | Present Orbital Ranks $n$ | $\Delta n$ | Ref. vsin$i$ & $P_{rot}$ | B-V | Rotation Period $P_{rot}$ at 650 Myr | Orbital Ranks $n$ at 650 Myr | $\Delta n$ |
|---|---|---|---|---|---|---|---|---|---|---|---|
| HD 59686b | 303.00 | 11.62 | 4.28 | 137.32 | 1.30 | -0.20 | (32) | 1.126 | 11.92 | 2.94 | -0.06 |
| HD 60532 b | 201.30 | 3.20 | 8.00 | 20.23 | 2.15 | 0.15 | (75) | 0.520 | 2.50 | 4.32 | -0.18 |
| HD 60532 c | 607.30 | 3.20 | 8.00 | 20.23 | 3.11 | 0.11 | (75) | 0.520 | 2.50 | 6.24 | 0.24 |
| HD 62509 b | 589.64 | | | 135.00 | 1.63 | 0.13 | (73) | 0.991 | 12.64 | 3.60 | 0.10 |
| HD 63454 b | 2.82 | | | 20.00 | 0.52 | 0.02 | (73) | | | | |
| HD 6434 b | 22.00 | | | 18.50 | 1.06 | 0.06 | (73) | 0.613 | 7.38 | 1.44 | -0.06 |
| HD 65216 b | 613.10 | 0.89 | 2.00 | 22.51 | 3.01 | 0.01 | (24) | 0.672 | 6.20 | 4.62 | 0.12 |
| HD 66428 b | 1,973.00 | 1.10 | 0.50 | 111.27 | 2.61 | 0.11 | (25) | 0.710 | 7.25 | 6.48 | -0.02 |
| HD 6718 b | 2,496.00 | 1.02 | 1.76 | 29.31 | 4.40 | -0.10 | (53) | 0.662 | 8.64 | 6.61 | 0.11 |
| HD 68988 b | 6.28 | | | 26.90 | 0.62 | 0.12 | (73) | 0.652 | 7.73 | 0.93 | -0.07 |
| HD 69830 b | 8.67 | | | 35.00 | 0.63 | 0.13 | (73) | 0.754 | 9.06 | 0.99 | -0.01 |
| HD 69830 c | 31.56 | | | 35.00 | 0.97 | -0.03 | (73) | 0.754 | 9.06 | 1.52 | 0.02 |
| HD 69830 d | 197.00 | | | 35.00 | 1.78 | -0.22 | (73) | 0.754 | 9.06 | 2.79 | -0.21 |
| HD 70573 b | 851.80 | | | 3.30 | 6.37 | -0.13 | (73) | 0.590 | | | |
| HD 70642 b | 2,231 | | | 31.17 | 4.15 | 0.15 | (73) | 0.677 | 6.20 | 7.11 | 0.11 |
| HD 72659 b | 3,177.40 | | | 20.50 | 5.37 | -0.13 | (73) | 0.612 | 7.38 | 7.55 | 0.05 |
| HD 73256 b | 2.55 | | | 13.97 | 0.57 | 0.07 | (73) | 0.781 | 12.04 | 0.60 | 0.10 |
| HD 73267 b | 1,260 | | | 42.00 | 3.11 | 0.11 | (51) | 0.806 | 5.49 | 6.12 | 0.12 |
| HD 73526 b | 188.30 | | | 36.44 | 1.73 | 0.23 | (73) | 0.737 | 9.13 | 2.74 | 0.24 |
| HD 73526 c | 377.80 | | | 36.44 | 2.18 | 0.18 | (73) | 0.737 | 9.13 | 3.46 | -0.04 |
| HD 73534 b | 1,800 | 2.65 | 0.50 | 268.07 | 1.89 | -0.11 | (75) | 0.962 | 9.36 | 5.77 | -0.23 |
| HD 74156 b | 51.64 | | | 19.00 | 1.40 | -0.10 | (73) | 0.585 | 5.64 | 2.09 | 0.09 |
| HD 74156 c | 2,476 | | | 19.00 | 5.07 | 0.07 | (73) | 0.585 | 5.64 | 7.60 | 0.10 |
| HD 75289 b | 3.51 | | | 16.83 | 0.59 | 0.09 | (73) | 0.578 | 4.72 | 0.91 | -0.09 |
| HD 75898 b | 204.20 | 1.71 | 4.00 | 21.56 | 2.12 | 0.12 | (24) | 0.630 | 8.65 | 2.87 | -0.13 |
| HD 76700 b | 3.97 | | | 31.77 | 0.50 | 0.00 | (73) | 0.745 | 9.06 | 0.76 | -0.24 |
| HD 7924 b | 5.40 | 0.78 | 1.35 | 29.22 | 0.57 | 0.07 | (75) | 0.826 | 9.71 | 0.82 | -0.18 |
| HD 80606 b | 111.44 | | | 41.00 | 1.40 | -0.10 | (73) | 0.765 | 7.15 | 2.50 | 0.00 |
| HD 81040 b | 1,001.70 | | | 9.80 | 4.68 | 0.18 | (73) | 0.680 | 6.20 | 5.45 | -0.05 |
| HD 81688 b | 184.02 | 13.00 | 1.10 | 597.75 | 0.68 | 0.18 | (24) | 0.993 | 12.64 | 2.44 | -0.06 |
| HD 82943 b | 441.20 | | | 18.00 | 2.90 | -0.10 | (46) | 0.620 | 7.38 | 3.91 | -0.11 |
| HD 82943 c | 219.40 | | | 18.00 | 2.30 | -0.20 | (46) | 0.620 | 7.38 | 3.10 | 0.10 |
| HD 83443 b | 2.99 | | | 35.00 | 0.44 | -0.06 | (73) | 0.811 | 5.49 | 0.82 | -0.18 |
| HD 8535 b | 1,313 | 1.19 | 1.41 | 42.69 | 3.13 | 0.13 | (53) | 0.553 | 3.00 | 7.59 | 0.09 |
| HD 85390 b | 788.00 | | | 44.00 | 2.62 | 0.12 | (50) | 0.860 | 9.18 | 4.41 | -0.09 |
| HD 8574 b | 227.55 | | | 18.00 | 2.33 | -0.17 | (73) | 0.580 | 5.64 | 3.43 | -0.07 |
| HD 86081 b | 2.14 | | | 24.83 | 0.44 | -0.06 | (73) | 0.664 | 8.64 | 0.63 | 0.13 |
| HD 86264 b | 1,475 | 1.66 | 12.50 | 6.72 | 6.03 | 0.03 | (75) | 0.460 | 2.90 | 7.98 | -0.02 |
| HD 8673 b | 639.00 | 1.45 | 30.00 | 2.44 | 6.40 | -0.10 | (24) | 0.489 | 3.70 | 5.57 | 0.07 |
| HD 87883 b | 2,754 | 0.76 | 2.17 | 17.71 | 5.38 | -0.12 | (75) | 0.960 | 9.36 | 6.65 | 0.15 |
| HD 88133 b | 3.42 | | | 49.91 | 0.41 | -0.09 | (73) | 0.810 | 5.49 | 0.85 | -0.15 |
| HD 89307 b | 3,090.00 | | | 18.00 | 5.56 | 0.06 | (73) | 0.594 | 5.64 | 8.18 | 0.18 |
| HD 89744 b | 256.00 | | | 9.00 | 3.05 | 0.05 | (73) | 0.530 | 5.87 | 3.52 | 0.02 |
| HD 90156 b | 49.77 | | | 26.00 | 1.24 | 0.24 | (50) | 0.660 | 7.43 | 1.89 | -0.11 |
| HD 92788 b | 325.81 | | | 21.30 | 2.48 | -0.02 | (46) | 0.694 | 7.25 | 3.56 | 0.06 |
| HD 93083 b | 143.58 | | | 48.00 | 1.44 | -0.06 | (73) | 0.940 | 9.36 | 2.48 | -0.02 |
| HD 9446 b | 30.05 | 1.00 | 4.00 | 12.64 | 1.33 | -0.17 | (75) | 0.690 | 7.25 | 1.61 | 0.11 |
| HD 9446 c | 192.90 | 1.00 | 4.00 | 12.64 | 2.48 | -0.02 | (75) | 0.690 | 7.25 | 2.99 | 0.00 |
| HD 95089 b | 507.00 | 4.90 | 2.74 | 90.45 | 1.78 | -0.22 | (40) | 0.940 | 9.36 | 3.78 | -0.22 |
| HD 96167 b | 498.90 | 1.86 | 3.80 | 24.76 | 2.72 | 0.22 | (75) | 0.730 | 7.15 | 4.12 | 0.12 |
| HD 97658 b | 9.49 | 0.73 | 0.50 | 73.84 | 0.50 | 0.00 | (73) | 0.795 | 12.04 | 0.92 | -0.08 |
| HD 99109 b | 439.30 | | | 48.00 | 2.09 | 0.09 | (73) | 0.867 | 9.18 | 3.63 | 0.13 |
| HD 99492 b | 17.04 | 0.81 | 2.30 | 17.81 | 0.99 | -0.01 | (16) | 1.000 | 12.64 | 1.10 | 0.10 |
| HD 99492 c | 4,970 | 0.81 | 1.36 | 30.12 | 5.48 | -0.02 | (16) | 1.000 | 12.64 | 7.33 | -0.17 |
| HD 102956 b | 6.50 | 4.40 | 0.30 | 741.82 | 0.21 | 0.21 | (40) | 0.971 | 12.64 | 0.80 | -0.20 |
| HD 1461 b | 5.77 | 1.10 | 1.60 | 34.61 | 0.55 | 0.05 | (61) | 0.697 | 7.25 | 0.93 | -0.07 |

| Exoplanet | Planet Orbital Period $P_n$ | Stellar Radius (Rs) | $v \sin i$ (km s⁻¹) | Present Rotation Period $P_{rot}$ | Present Orbital Ranks $n$ | $\Delta n$ | Ref. $v\sin i$ & $P_{rot}$ | B-V | Rotation Period $P_{rot}$ at 650 Myr | Orbital Ranks $n$ at 650 Myr | $\Delta n$ |
|---|---|---|---|---|---|---|---|---|---|---|---|
| HD 156668 b | 4.65 | | | 51.50 | 0.45 | -0.05 | (36) | 1.015 | 10.82 | 0.75 | 0.25 |
| HIP 14810 b | 6.67 | 1.00 | 1.50 | 33.72 | 0.58 | 0.08 | (25) | 0.777 | 5.61 | 1.06 | 0.06 |
| HIP 14810 c | 95.29 | 1.00 | 1.50 | 33.72 | 1.41 | -0.09 | (25) | 0.777 | 5.61 | 2.57 | 0.07 |
| HIP 14810 d | 962.00 | 1.00 | 1.50 | 33.72 | 3.06 | 0.06 | (25) | 0.777 | 5.61 | 5.56 | 0.06 |
| HIP 5158 b | 345.72 | 0.71 | 1.57 | 22.87 | 2.47 | -0.03 | (75) | 1.078 | 10.82 | 3.17 | 0.17 |
| HIP 57050 b | 41.40 | | | 98.10 | 0.75 | -0.25 | (27) | 1.600 | 0.68 | 3.93 | -0.07 |
| kappa CrB b | 1,191 | 4.71 | 3.00 | 79.41 | 2.47 | -0.03 | (16) | 1.002 | 12.64 | 4.55 | 0.05 |
| Kepler-4 b | 3.21 | 1.49 | 2.10 | 35.81 | 0.45 | -0.05 | (75) | 0.390 | 1.67 | 1.24 | -0.26 |
| Kepler-5 b | 3.55 | 1.79 | 4.80 | 18.89 | 0.57 | 0.07 | (75) | 0.409 | 1.67 | 1.29 | -0.21 |
| Kepler-6 b | 3.23 | 1.39 | 3.00 | 23.45 | 0.52 | 0.02 | (75) | 1.410 | 3.66 | 0.96 | -0.04 |
| Kepler-7 b | 4.89 | 1.84 | 4.20 | 22.19 | 0.60 | 0.10 | (75) | 0.370 | 1.67 | 1.43 | -0.07 |
| Kepler-8 b | 3.52 | 1.49 | 10.50 | 7.16 | 0.79 | -0.21 | (75) | | | | |
| OGLE2-TRL9 | 2.49 | 1.53 | 39.33 | 1.97 | 1.08 | 0.08 | (66) | | | | |
| OGLE-TR10 | 3.10 | | | 15.80 | 0.58 | 0.08 | (73) | 0.606 | 5.10 | 0.85 | -0.15 |
| OGLE-TR111 | 4.01 | | | 38.00 | 0.47 | -0.03 | (73) | 0.933 | 3.66 | 1.03 | 0.03 |
| OGLE-TR113 | 1.43 | | | 31.10 | 0.36 | -0.14 | (73) | 0.340 | 1.67 | 0.95 | -0.05 |
| OGLE-TR132 | 1.69 | 1.34 | 5.00 | 13.56 | 0.50 | 0.00 | (49) | 0.820 | 9.71 | 0.56 | 0.06 |
| OGLE-TR-56 | 1.21 | | | 26.30 | 0.36 | -0.14 | (73) | 0.589 | 5.64 | 0.60 | 0.10 |
| Qatar-1 b | 1.42 | 0.82 | 2.10 | 19.82 | 0.42 | -0.08 | (2) | | | | |
| rho CrB b | 39.65 | | | 17.00 | 1.33 | -0.17 | (73) | 0.608 | 5.10 | 1.98 | -0.02 |
| Tau Boo b | 3.31 | | | 4.00 | 0.94 | -0.06 | (7) | 0.498 | 3.70 | 0.96 | -0.04 |
| TrES-1 | 3.03 | | | 34.77 | 0.44 | -0.06 | (73) | 0.922 | 10.26 | 0.67 | 0.17 |
| TrES-2 | 2.47 | | | 24.78 | 0.46 | -0.04 | (73) | 0.923 | 10.26 | 0.62 | 0.12 |
| TrES-3 | 1.31 | 0.80 | 2.00 | 20.28 | 0.40 | -0.10 | (57) | 0.592 | 5.64 | 0.61 | 0.11 |
| TrES-4 | 3.55 | 1.81 | 9.50 | 9.64 | 0.72 | 0.22 | (45) | 0.436 | 2.90 | 1.07 | 0.07 |
| Ups And b | 4.62 | 1.63 | 9.62 | 8.58 | 0.81 | -0.19 | (16) | 0.496 | 3.70 | 1.08 | 0.08 |
| Ups And c | 237.70 | 1.63 | 9.62 | 8.58 | 3.03 | 0.03 | (16) | 0.496 | 3.70 | 4.01 | 0.01 |
| Ups And d | 1,302.61 | 1.63 | 9.62 | 8.58 | 5.34 | -0.16 | (16) | 0.496 | 3.70 | 7.06 | 0.09 |
| Ups And e | 3,848.86 | 1.63 | 9.62 | 8.58 | 7.66 | 0.16 | (16) | 0.496 | 3.70 | 10.13 | 0.13 |
| WASP-1 b | 2.52 | 1.38 | 5.77 | 12.11 | 0.59 | 0.09 | (75) | 0.200 | 1.67 | 1.15 | 0.15 |
| WASP-10 b | 3.09 | 0.78 | 3.00 | 13.20 | 0.62 | 0.12 | (75) | 0.570 | 4.72 | 0.87 | -0.13 |
| WASP-11 b | 3.72 | 0.81 | 0.50 | 81.94 | 0.36 | -0.14 | (75) | 1.000 | 12.64 | 0.67 | 0.17 |
| WASP-12 b | 1.09 | 1.57 | 2.20 | 36.09 | 0.31 | -0.19 | (75) | 0.420 | 1.67 | 0.87 | -0.13 |
| WASP-13 b | 4.35 | 1.34 | 4.90 | 13.83 | 0.68 | 0.18 | (75) | 0.890 | 10.26 | 0.75 | 0.25 |
| WASP-14 b | 2.24 | 1.30 | 4.90 | 13.39 | 0.55 | 0.05 | (75) | 0.460 | 2.90 | 0.92 | -0.08 |
| WASP-15 b | 3.75 | 1.48 | 4.00 | 18.68 | 0.59 | 0.09 | (75) | 0.389 | 1.67 | 1.31 | -0.19 |
| WASP-16 b | 3.12 | 0.95 | 3.00 | 15.95 | 0.58 | 0.08 | (75) | 1.220 | 14.14 | 0.60 | 0.10 |
| WASP-17 b | 3.74 | 1.38 | 9.00 | 7.76 | 0.78 | -0.22 | (75) | 0.240 | 1.67 | 1.31 | -0.19 |
| WASP-18 b | 0.94 | 1.23 | 11.00 | 5.66 | 0.55 | 0.05 | (75) | 0.490 | 3.70 | 0.63 | 0.13 |
| WASP-19 b | 0.79 | 0.93 | 4.00 | 11.76 | 0.41 | -0.09 | (75) | 0.330 | 1.67 | 0.78 | -0.22 |
| WASP-21 b | 4.32 | 1.06 | 1.50 | 35.74 | 0.49 | -0.01 | (14) | 0.200 | 1.67 | 1.37 | -0.13 |
| WASP-22 b | 3.53 | 1.13 | 3.50 | 16.33 | 0.60 | 0.10 | (75) | 0.427 | 1.67 | 1.28 | -0.22 |
| WASP-24 b | 2.34 | 1.15 | 6.96 | 8.34 | 0.65 | 0.15 | (67) | 0.745 | 9.06 | 0.64 | 0.14 |
| WASP-25 b | 3.76 | 0.95 | 3.00 | 16.02 | 0.62 | 0.12 | (21) | 0.415 | 1.67 | 1.31 | -0.19 |
| WASP-26 b | 2.76 | 1.34 | 2.40 | 28.24 | 0.46 | -0.04 | (75) | 0.345 | 1.67 | 1.18 | 0.18 |
| WASP-28 b | 3.41 | 1.05 | 2.20 | 24.14 | 0.52 | 0.02 | (74) | 0.470 | 2.80 | 1.07 | 0.07 |
| WASP-29 b | 3.92 | 0.85 | 1.50 | 28.53 | 0.52 | 0.02 | (33) | 0.818 | 5.49 | 0.89 | -0.11 |
| WASP-3 b | 1.85 | 1.31 | 14.10 | 4.70 | 0.73 | 0.23 | (75) | 0.370 | 1.67 | 1.03 | 0.03 |
| WASP-31 b | 3.41 | 1.12 | 7.90 | 7.17 | 0.78 | -0.22 | (75) | 0.368 | 1.67 | 1.27 | 0.27 |
| WASP-32 b | 2.72 | 1.11 | 4.80 | 11.70 | 0.61 | 0.11 | (75) | 0.798 | 12.04 | 0.61 | 0.11 |
| WASP-33 b | 1.22 | 1.44 | 90.00 | 0.81 | 1.15 | 0.15 | (17) | 0.700 | 7.25 | 0.55 | 0.05 |
| WASP-34 b | 4.32 | 0.93 | 1.40 | 33.60 | 0.50 | 0.00 | (75) | 0.770 | 5.61 | 0.92 | -0.08 |
| WASP-37 b | 3.58 | 0.98 | 2.40 | 20.59 | 0.56 | 0.06 | (65) | 0.600 | 5.10 | 0.89 | -0.11 |
| WASP-38 b | 6.87 | 1.37 | 8.58 | 8.05 | 0.95 | -0.05 | (75) | | | | |
| WASP-4 b | 1.34 | 1.15 | 2.20 | 26.44 | 0.37 | -0.13 | (75) | 1.026 | 10.82 | 0.50 | 0.00 |
| WASP-5 b | 1.63 | 1.08 | 3.40 | 16.13 | 0.47 | -0.03 | (75) | 0.480 | 2.80 | 0.83 | -0.17 |

| Exoplanet | Planet Orbital Period $P_n$ | Stellar Radius (Rs) | $v \sin i$ (km s$^{-1}$) | Present Rotation Period $P_{rot}$ | Present Orbital Ranks $n$ | $\Delta n$ | Ref. $v\sin i$ & $P_{rot}$ | B-V | Rotation Period $P_{rot}$ at 650 Myr | Orbital Ranks $n$ at 650 Myr | $\Delta n$ |
|---|---|---|---|---|---|---|---|---|---|---|---|
| WASP-6 b | 3.36 | 0.87 | 1.40 | 31.43 | 0.47 | -0.03 | (75) | 0.880 | 10.26 | 0.69 | 0.19 |
| WASP-7 b | 4.95 | 1.24 | 17.00 | 3.68 | 1.10 | 0.10 | (75) | 1.442 | 3.66 | 1.11 | 0.11 |
| WASP-8 b | 8.16 | 0.95 | 1.58 | 30.51 | 0.64 | 0.14 | (60) | | | | |
| xi Aql b | 136.75 | 12.00 | 19.00 | 31.94 | 1.62 | 0.12 | (35) | 1.023 | 10.82 | 2.33 | -0.17 |
| XO-1 b | 3.94 | 1.18 | 1.11 | 53.95 | 0.42 | -0.08 | (48) | 0.577 | 4.72 | 0.94 | -0.06 |
| XO-2 b | 2.62 | 0.96 | 1.40 | 34.83 | 0.42 | -0.08 | (16) | 0.820 | 9.71 | 0.65 | 0.15 |
| XO-3 b | 3.19 | 1.38 | 18.54 | 3.76 | 0.95 | -0.05 | (39) | 0.450 | 2.90 | 1.03 | 0.03 |
| XO-5 b | 4.19 | 1.06 | 1.80 | 29.79 | 0.52 | 0.02 | (75) | 0.840 | 9.18 | 0.77 | -0.23 |

## 5. STATISTICAL ANALYSIS OF QUANTIZATION RESULTS

### 5.1 Statistical Test for $\Delta n$ deviations from half-integer values

Form Sect. 4.1, the distribution of the orbital ranks $n$ for the 443 exoplanets (in bins of 0.1 increment) that were calculated for the present stellar rotation periods is presented in Fig. 1, along with the distribution of deviations $\Delta n$ from half-integers in Fig. 2. It can be observed that the orbital ranks $n$ indeed tend to cluster around integer or half-integer values, similar to the Solar System results. Around 288 exoplanets (65 per cent) have absolute deviations of $|\Delta n| < 0.1$. If there is no correlation between the stellar rotation periods and the planetary orbital periods, the deviations from half-integer $\Delta n$ would follow a uniform distribution as opposed to the distribution observed in Fig. 2. To test that the results were not obtained by chance, we therefore need to reject the null hypothesis corresponding to a uniform distribution of deviations $\Delta n$ in the interval [-0.25, 0.25]. A Kolmogorov-Smirnov one-sample test between the observed cumulative $\Delta n$ distribution and that of a uniform distribution yields a maximum difference statistic $D=0.54$ and a probability $p<0.046$ to have been obtained by chance. Using a $\chi^2$ test with 20 bins, we obtain $\chi^2=32.89$ and the probability for $\Delta n$ to be drawn from a uniform distribution, i.e. the probability for the results to be obtained by chance is $p<0.024$.

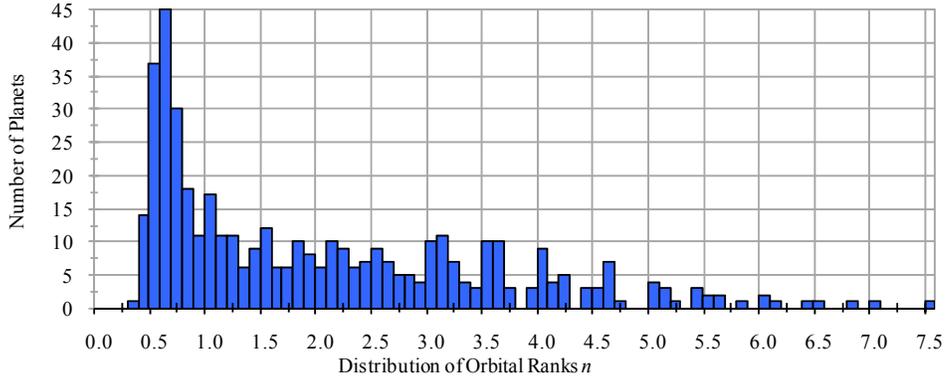

**Figure 1** – Distribution of present orbital ranks *n* for the 443 exoplanets indicating an obvious clustering around discrete half-integer values, with peaks at *n*=0.5, 1.0, 1.5, 2.0, 3.0, 3.5, 4.0, 4.5, 5.0, 6.0

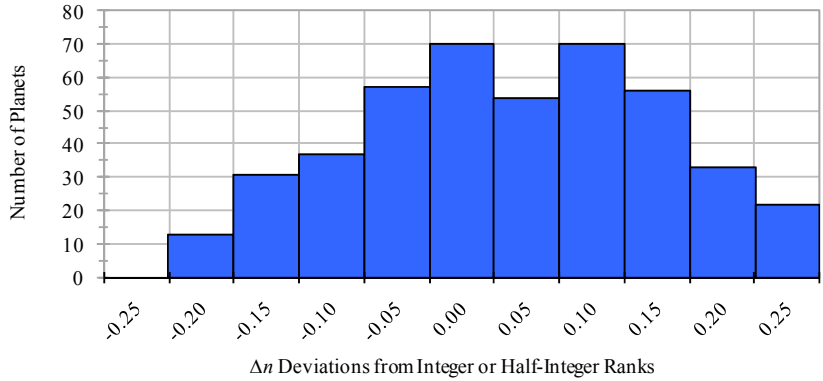

**Figure 2** – Distribution of Δ*n* deviations from the nearest half-integer for the 443 exoplanets, indicating that 65% of deviations are within +/- 0.1 of the nearest half-integer. The probability to obtain these results by chance is p<0.024.

### 5.2 Monte-Carlo Simulation for uncertainty in present vsini and rotation periods

To address the uncertainty in rotation periods derived from stellar radii and *v*sin*i*, we have used a Monte-Carlo treatment and calculated the orbital ranks from 25 randomly generated rotation periods ranging within +/- 20 per cent of the estimated rotation period for each of the 443 exoplanetary stars and resulting in a total of 11,075 simulated orbital ranks. The Monte-Carlo distribution of orbital ranks is presented in Fig.3 and their deviation from half-integer Δ*n* is presented in Fig.4. We note a clustering of orbital ranks around half-integers for *n*=0.5, 1.0, and 1.5, representing around 50 percent of all exoplanets, while the distribution of orbital ranks beyond *n*>2.0 starts to resemble a uniform distribution. This is expected since the uncertainty in rotation periods discussed in Sect. 4.1, was expected to result in a 10 percent uncertainty in orbital ranks *n*, which in turn exceeds the critical absolute deviation |Δ*n*|=0.25 for orbital ranks higher than *n*>2.0. Nonetheless, a Kolmogorov-Smirnov one-sample test between the observed cumulative Δ*n* distribution and that of a uniform distribution yields a maximum difference statistic *D*=0.54 and a probability *p*<0.046 to have been obtained by chance.

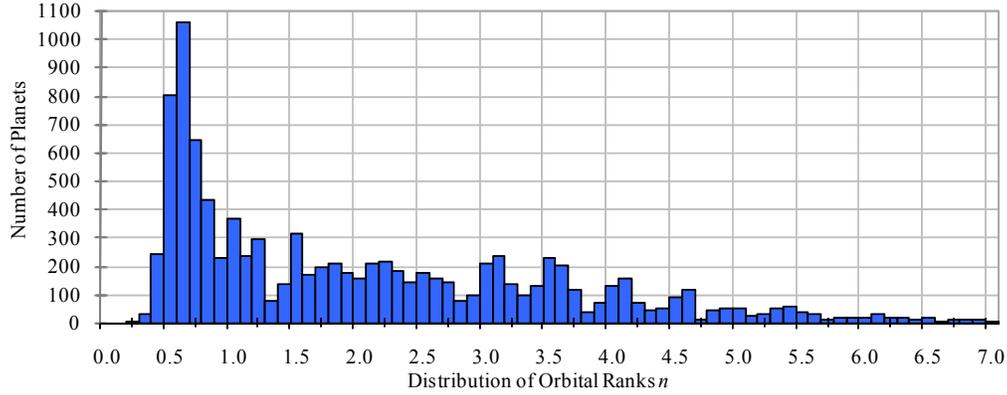

**Figure 3** – Distribution of orbital ranks *n* calculated from 25 randomly generated rotation periods (within 20% uncertainty) for each of the 443 exoplanets, indicating a clustering around discrete half-integer values with obvious peaks at n=0.5, 1, 1.5, 3.0, 3.5, 4.0, 4.5, 5.0.

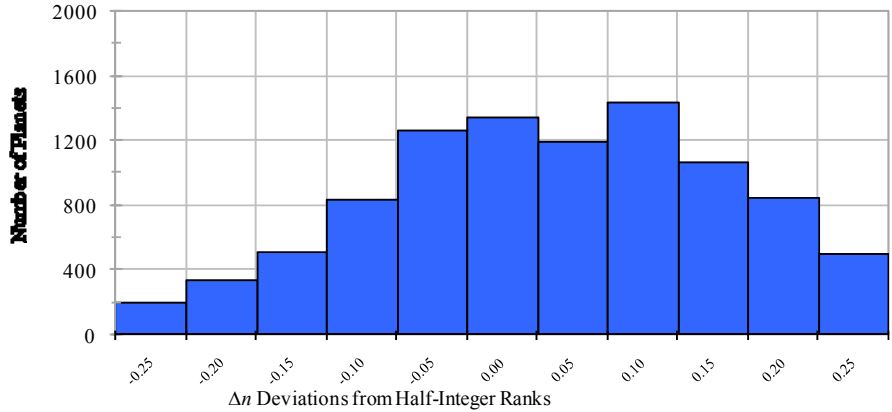

**Figure 4** – Distribution of deviations from half-integers Δ*n* calculated from the 25 randomly generated rotation periods (within 20% uncertainty) for each of the 443 exoplanetary parent stars. The probability to obtain these results by chance is *p*<0.046.

### 5.3 Statistical Test for Δn distribution using estimated rotation periods at formation age

Using the alternative approach of Sect. 4.2, the distribution of orbital ranks calculated from the stellar rotation periods at the formation epoch, which were derived from the B-V color of Hyades stars at the fiducial age of 650 Myr, is presented in Fig. 5 and the distribution of deviations from half-integer is presented in Fig.6. Clustering around half-integers is again apparent. If the results were obtained by chance, the deviations from half-integer Δ*n* would follow a uniform distribution. However, out of the 443 exoplanets sample, 252 planets (57%) have |Δ*n*|<0.1. The mean absolute deviation from integer or half-integer is |Δ*n*|=0.09 with a standard deviation of 0.05. The average deviations taken as a percentage of the orbital ranks, Δ*n*/*n* is 5 per cent. A Kolmogorov-Smirnov one-sample test between the observed cumulative Δ*n* distribution and that of a uniform distribution yields a maximum difference statistic *D*=0.54. For the 443 data points, this result has a probability *p*<0.046 to be obtained by chance. Using a $\chi^2$ test with 11 degrees of freedom (bins), we obtain $\chi^2$=57.4 and the probability for Δ*n* to be drawn from a uniform distribution, i.e. the probability for the results to be obtained by chance is *p*<2.8x10$^{-8}$.

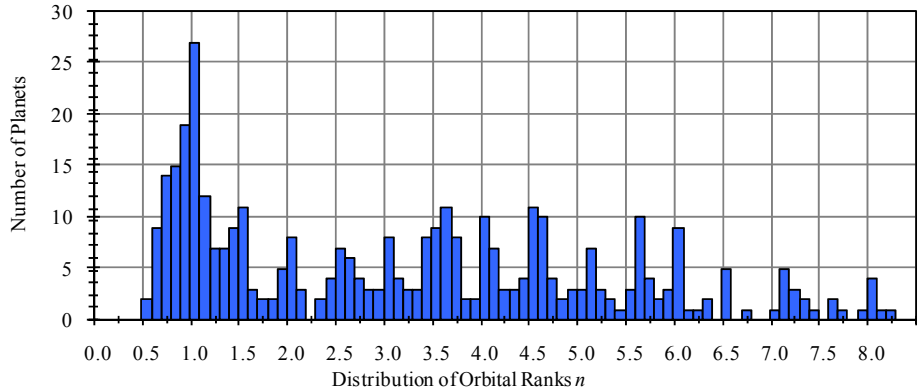

**Figure 5** – Distribution of orbital ranks *n* for the 443 exoplanets, calculated from rotation periods of matching Hyades stars at the fiducial age of 650 Myr, again indicating an obvious clustering around discrete integer or half-integer values.

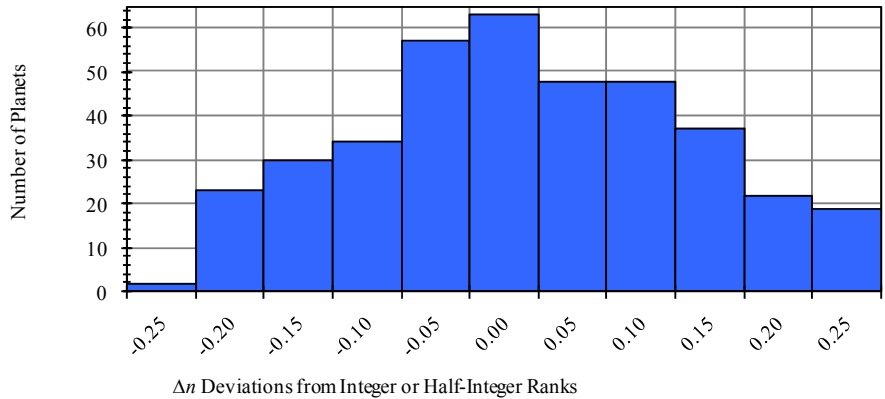

**Figure 6** – Distribution of Δ*n* deviations from the nearest half-integer for the 443 exoplanets, calculated using the matching Hyades stars at formation age of 650 Myr. The probability to obtain these results by chance is p<2.8x10$^{-8}$.

### 5.4 Monte-Carlo Simulation for uncertainty in estimated rotation periods at formation age

According to Radick et al. (1995), the measured rotation periods for Hyades stars vary between 2 to 8 per cent from year to year and have an accuracy of +/- 0.1 day. We have taken the worst case and considered 8 per cent uncertainty in all measured rotation periods, which translates to around +/- 1 day for the majority of the stellar rotation periods. To address this inherent uncertainty, we apply a Monte-Carlo treatment and calculate the orbital ranks *n,* using 50 randomly generated rotation periods (within +/- 8 per cent of the measured rotation periods) for each of the 443 exoplanetary stars, resulting in a total of 11,075 orbital ranks. The Monte-Carlo distribution of orbital ranks and their deviation from half-integer Δ*n* is presented in Fig.4. Again, a clustering around half-integers is obvious, all the way up to orbital ranks *n*<6, representing around 90% of all exoplanets. A Kolmogorov-Smirnov one-sample test yields a maximum difference statistic *D*=0.52 between the observed cumulative distribution and that of a uniform distribution. For the 11,075 data points, this result has a probability

*p*<0.0035 to be obtained by chance. Using a $\chi^2$ test with 20 degrees of freedom (bins), we obtain $\chi^2$=1,835 and the probability for the results to be obtained by chance is nil.

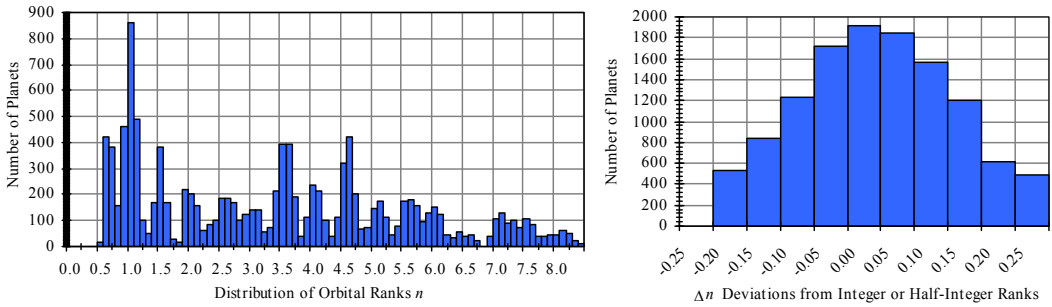

**Figure 7** – Distribution of orbital ranks *n* calculated from 50 randomly generated rotation periods (within 8% uncertainty) for each of the 443 exoplanets, indicating an obvious clustering around discrete integer or half-integer values. The probability to obtain these results by chance is almost nil.

### 5.5 Test for mathematical artifacts

In order to investigate the origin of the apparent quantization in Fig. 1 to 7, we shuffled the stellar and planetary angular velocities, to give randomized matches between the properties of the stars and the properties of the planets, to see if the quantization persists. As a result, the Δ*n* deviations gave a uniform distribution in the interval [-0.25, 0.25], and the quantized distribution disappeared, which again supports the fact that quantization hypothesis did not occur by chance.

It is important to note that while the obtained quantization results are valid for stellar rotation periods both at present and at the formation age, it is observed that the tendency to cluster around half-integers seems to improve and become more pronounced as stellar rotation P$_{rot}$ increases and rotation slows down. The mean absolute |Δ*n*| deviation from half-integer is 0.1 for rotation periods derived at the present age and is also the same value for those derived at the formation age. However, the standard deviation for |Δ*n*| derived from present rotation periods is 0.068 and is improved compared to the standard deviation of 0.075 for |Δ*n*| derived from estimated rotation periods at formation age. Additionally, the |Δ*n*| distribution's peak at formation age is more flat (with a positive Kutosis *k*=0.97) and becomes much sharper (with a negative Kurtosis *k*= - 0.71) as rotation slows down to the present values, i.e. the quantization features improve.

The distribution of the ratio *r* of the planetary orbital period to the stellar rotation period is not uniform, but decreases with increasing *r* due to the bias in detecting exoplanets with short orbital periods. In order to test if the above quantization results is fundamental in nature or whether it is simply the result of this selection effect, the following test was proposed. A continuous 'toy' distribution of *r* is used which has the gross properties of the actual distribution that is, one which reflects the observational biases rather than real biases. A histogram for the distribution of orbital ranks *n* is then generated using the same method as in Section 4, i.e. using the cubic root relationship of Eq. (2.9). We then searched to see if any quantization features persist, which would then support the hypothesis that they are not fundamental in nature, but effectively a mathematical artifact. However, as shown in Fig. 8, the quantization features almost disappeared and no major clustering around half-integer was observed except for a peak at 0.6 which was expected due to the large number of short-period planets observed. That distribution around the peak at *n*=0.6 is gradual, uniform, and not as sharp when compared to the distribution around *n*=0.5 in Fig.1 and Fig 3. Moreover, the original peaks in Fig. 1 and Fig. 5 at *n*=1.0, 1.5, 2.0, 3.0, and 3.5 have almost disappeared in

Fig. 8. This implies that the hypothesized quantization is not the result of mathematical artifacts and must have some fundamental physical basis to it.

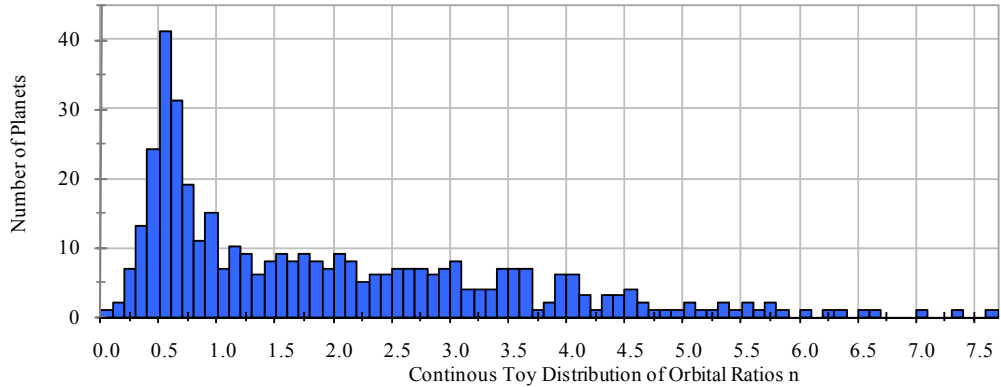

**Figure 8** – Distribution of orbital ranks *n* calculated from a 'toy' distribution of *r* (*r* being the ratio of planetary orbital period to the star's rotation period) and shows insignificant quantization features, supporting that the hypothesized quantization is not the result of mathematical artifacts.

### 5.6 Statistical Analysis Summary

In summary, if the planetary orbital periods are not quantized with respect to the stellar rotation period, the orbital ranks *n* and their deviation $\Delta n$ from half-integer values would have exhibited a uniform distribution. However, the peaks in orbital ranks *n* at half-integer and the observed distribution in $\Delta n$, indicates that this is not the case and that the probability to obtain such a distribution by chance is *p*<0.024. This was done for orbital ranks derived both from present rotation periods and from those estimated at the formation age (~ 650 Myr).

The uncertainties in *v*sin*i* and rotation periods were dealt with using a Monte-Carlo treatment by generating random rotation periods within a range of 20 per cent of the estimated rotation periods. The derived distribution of orbital ranks continued to show some quantization peaks, which were however less pronounced. The test for mathematical artifacts proves that a uniform distribution of *r*, i.e. the ratio of planetary orbital periods to the star's rotation period, does not produce the hypothesized quantization results.

In the above statistical analysis, we did not include the Solar System results, for which the planetary orbital periods and the Sun's rotation period are accurately well determined. Regardless of the uncertainties in stellar rotation periods, the orbital ranks *n* in the Solar System are clearly quantized over half-integer values and this provides additional support to the quantization hypothesis.

## 6. DISCUSSION

### 6.1 Quantized Orbits Featuring in Hybrid models of Planet Formation

The statistically significant results confirm the hypothesis that the specific orbital angular momenta of planets in the solar and extra solar systems tend to be discrete and quantized, clustering around half-integer multiples of the specific angular momentum at the central star's corotation radius. This half-integer orbital quantization is therefore directly dependent on the parent star's rotation rate and is not related to any universal constant, but is a system-specific physical property. This provides a more physical description of the Titus-Bode empirical law featured in the Solar System and possibly some extra-solar multi-planetary systems, which can now be interpreted using a quasi-quantum physical model with stellar rotation as the main quantization parameter.

One possible theoretical justification for having half-integers as the fundamental unit of quantization and not some other fraction, such 1/3, 1/5, or even integer multiples, might have to do with the similarity of the planetary quantization with Bohr-Sommerfeld's atomic quantum model. The planetary quantization and its dependency on stellar rotation is consistent with a corresponding dependency at the atomic scale, where the discrete angular momenta of electron orbits are integer multiples of Planck's reduced constant $\hbar$. However, the electrons orbits are similarly related to the spin angular momentum of the system's 'central body' (the proton in the Hydrogen atom) whose value is ½ $\hbar$. The atomic unit of spin quantization is similarly in multiples of ½ $\hbar$, which by using Bohr's 'Correspondence Principle' is consistent with the half-integer planetary quantization obtained here. Regardless of the theoretical justification, the study this planetary quantization with it dependency on stellar rotation is better explained as a product of physical mechanisms involved in the planets formation process.

The diverse properties of exoplanets, i.e. semi-major axes, orbital eccentricities, masses, and inclinations, are the result of a combination of events that took place during the initial formation stage (which includes possible migration and disc interactions), and the longer-term dynamical evolution stage which followed after the protoplanetary disc dissipated. It is not yet clear which of the two stages govern the shaping of the system's dynamical properties and long term stability. Both current theories of planet formation, core accretion and gravitation instability have their limitations (D'Angelo et al 2010).  The core-accretion model suffering from time-scales that are too long for observed disc lifetimes and the gravitational instability model having some difficulties explaining the low disc temperatures needed for its operation. A new trend of hybrid models has emerged where the virtues of both models complement each other. In such hybrid models, (e.g. Durisen et al. 2005) concentric dense gas rings created by gravitational instabilities enhance the growth rate of solid cores by drawing solids toward their centers, thus accelerating core-accretion and runaway growth (Haghighipour & Boss 2003). Durisen et al. (2005) indicated that the dense rings appear to be produced by resonances with discrete spiral modes which we suggest can be correlated with a quantum-like structure. The hybrid gravitational instability model of planet formation appears to be the most suited for explaining the reported quantization of planetary orbits. One reason is that the gravitational instability model has been successfully used in the past to explain 'discrete' power law distributions such as the mathematical regularity in planetary spacing observed in the empirical Titius-Bode law (Griv & Gedalin 2005). Another reason is that it requires minimal orbital migration, at least initially, because the self-gravitating disc gas flows inward, past the protoplanets, leaving them relatively undisturbed (e.g. Boss 2005). This implies that planets can form directly in situ, or even by accelerated core-accretion, within the quantized spatial structures. However, even when protoplanets do migrate from their original birthplace, resonance can provide traps at discrete density jumps that enhance the accumulation of planetesimals (Masset et al. 2006) at these discrete 'quantized' locations.

*6.2 Disc Magnetospheric Truncation and the Lowest Ranking Orbits (at n=0.5)*

Various other natural mechanisms play a role in planetary structure and may explain the proposed quantization and its dependence on stellar rotation. One such possible mechanism is inner disc truncation by stellar magnetospheres (Lin & Papaloizou 1996). The stellar magnetic field of a spinning star couples to the protoplanetary disc and expels ionized gas from its innermost part, carving an inner gap at the truncation radius in the range of 3-10 stellar radii, depending on the disc accretion rate and magnetic field strength, which in turn is a function of the star's rotation rate. A migrating planet that reaches this inner gap can remain parked there indefinitely, no longer being dragged in by the accreting disc nor forced to exchange angular momentum with it. Hence, the truncation radius at the disc's inner gap serves as a physical inferior limit for planetary orbits and acts as the planets' last line of defense against their fall into the star. Yi (1995), modeled magnetic braking and found that the final size of the truncation gap was in the range of 3-10 stellar radii and that radius is highly dependent on the stellar rotation period. Therefore, the orbital angular momenta of migrating planets that end up at their parent star's truncation gap should correlate well with the star's rotation period.

From the half-integer quantization results of Table 3, the inferior limit of planetary orbits in any planetary system is at the discrete orbital rank $n$=0.5. In terms of the corotation radius $r_0$ and eq. (2.8), this inferior limit corresponds to a semi-major axis of $0.25r_0$. In order to verify the relationship between the lowest ranking orbits and the stars' disc truncation radii, we calculated the semi-major axis ($0.25r_0$) of the lowest ranking orbit (at $n$=0.5) for each exoplanetary system under consideration. We found that the lowest ranking orbits range from $0.02 - 0.09$ AU with a peak and mean semi-major axis of 0.043 AU ($\approx$ 8 mean stellar radii). The cumulative distribution, expressed in terms of the respective parent stars radii indicated that the lowest ranking orbits ($n$=0.5) are clustered in the range of 3-10 stellar radii, which is consistent with the predictions of protoplanetary disc models for magnetospheric inner gap sizes (e.g. Yi 1995). Hence, it is reasonable to suspect that the quantized lowest ranking orbits ($n$=0.5) are physically described by the parent stars disc's inner gap size at the magnetospheric truncation radius. This is significant, since the correlation between the disc's inner gap size and stellar rotation (Yi 1995) provides further support for the dependency of the quantized lowest ranking orbits on stellar rotation.

### 6.3 Tidal Dissipation and the Corotation Orbit (n=1.0)

Another mechanism that can explain the dependency of orbital ranks on stellar rotation is tidal dissipation. Lin et al. (1996) suggested that as a migrating planet approaches the central star, it will raise tidal bulges in that star which will transfer angular momentum from the rapidly spinning star to the more slowly spinning planet. The tidal dissipation within the star can circularize the planet's orbit and synchronize its orbital period to the star's rotation. The resulting spin-orbit coupling can be effective at pushing the planet outwards, keeping it at or near the corotation radius. Indeed, observations indicate that planets within 0.1 AU are nearly always on circular or nearly circular orbits, while beyond 0.3 AU the distribution of eccentricities appears essentially uniform between 0 and 0.8 (Butler et al. 2006). This observed split in the eccentricity–period distribution is evidence of orbital circularization for short-period planets by internal tidal dissipation (Rasio et al. 1996). The tidal locking mechanism can therefore provide a natural physical justification for exoplanets with discrete orbital ranks $n$=1, since by definition these planets are located at the corotation orbit. Recently, Alves et al. 2010 confirmed that the angular momentum of exoplanet parent stars follows, at least qualitatively, Kraft's relation. The mechanism of angular momentum transfer must then certainly have a role in producing this apparent quantization in the exoplanets angular momentum distribution.

### 6.4 Hot Jupiters & Multi-planetary Sample Analysis

The sample of 443 exoplanets can be split into 2 samples: 75 'hot Jupiters' and 368 non-hot Jupiters. We analyzed each sample separately to see if the quantization features would be different for systems with hot-Jupiters. In both samples, the distribution of deviations from half-integers remain almost the same with 65 percent of planets having an absolute $|\Delta n|$<0.1. However, we also found that more than 60 per cent of hot-Jupiters are clustered either at the orbital rank $n$=0.5 (which possibly represents the disc magnetospheric inner radius) or at $n$=1 (the corotation orbit), evidence of synchronization and spin-orbit coupling. Out of those, we found that the majority of hot-Jupiters with host-stars rotation of $P_{rot}$<10 days ($T_{eff} > 6000$ K) to be synchronized at the corotation orbit $n$=1, while the majority of hot-Jupiters with star rotation of $P_{rot}$>10 days ($T_{eff}$ < 6000 K) are orbiting even closer at $n$=0.5, i.e. at the proposed disc magnetospheric truncation radius.

We also examined the 49 multi-planetary systems in our samples, in which 6 systems harbor hot-Jupiter planets and found that the multi-planetary systems having hot-Jupiters tend to have on average a lower absolute deviation from half-integer (mean $|\Delta n|$=0.086) compared to multi-planetary systems that do not harbor any hot-Jupiters (mean $|\Delta n|$=0.107). This implies the quantization features may be more pronounced in multi-planetary systems with Hot-Jupiters. However, the sample is still too small to draw any solid conclusions at this stage and this will have to wait the discovery of more multi-planetary systems with hot-Jupiters.

### 6.5 The Role of Resonance Trapping in Forming Discrete Planetary Orbits

Resonance mechanisms may play an important role in explaining the quantized planetary orbits. Mean-motion resonances for instance have already been used to explain the sequence of planetary spacing in the empirical Titius-Bode's law (e.g. Patterson 1987). Additionally, mean-motion resonances were shown to influence the formation sites of protoplanets and were proposed as a means to halt planetary migration (resonance trapping). The migration of solid particles in a protoplanetary disc causes their orbits to decay and both eccentricity and inclination are damped with the loss of angular momentum. Under certain initial conditions however, resonance between the planetesimals and an already formed planet embryo can counteract this orbital decay and trap the particles in a stable resonant orbit. There are strong indications that this mechanism can also explain the near-commensurabilities of the Solar System outer planets (Beauge et al. 1994 and Malhotra 1995), as well as the spacing of the terrestrial planets (Laskar 1997).

Furthermore, resonance trapping was shown to be working in more than 20 per cent of the 19 multiple exoplanetary systems considered by Tinney et al. (2006). Motivated by these observations, different studies have shown that during migration the capture of giant planets into resonances is a natural expectation (Nelson & Papaloizou 2002). The planets subsequently migrate maintaining this commensurability. Resonance was also shown to play a key role in the formation of concentric density rings in the hybrid gravitational instability model of planet formation (Durisen et al. 2005). These resonant disc structures act as traps for infalling protoplanetary seeds and migrating planets and may provide a natural explanation for planetary orbits of higher discrete ranking ($n$>1). Similarly, within the context of multiple planets forming in a disc, migration of the innermost planet might be stopped by either the magnetospheric gap at the truncation radius ($n$=0.5) or by the star's tidal barrier at the corotation radius ($n$=1.0). The size of these inner orbits is highly dependent on the stellar rotation rate. Now a second protoplanet approaching the star would stop when entering a low order resonance with the innermost planet. The second planet's orbit would then be expected to correlate with the star's rotation rate as well. Therefore, the various features of resonance mechanisms may also provide a physical justification for the observed quantization in planetary orbits and their dependence on the star's rotation period. The multi-planetary system of Gl876 provides evidence for that. The inner planet Gl876d occupies the orbital rank of $n$=0.6 which we suggest may correspond to the disc magnetospheric truncation radius. The remaining planets Gl876 c, b, e are locked in a 1:2:4 Laplace resonance (with orbital ranks $n$= 1.58, 2.00, 2.54 respectively).

## 7. CONCLUSION & PROSPECTS

We have shown that the orbital structure of planetary systems exhibits quantized features that must have evolved from the dynamical process of planetary formation. Our results demonstrate that planetary orbital periods and the parent star rotation period are correlated by discrete integer or half-integer values. Of course, the orbital period quantization also implies a quantization of planetary angular momenta, semi-major axes, and mean orbital velocities as well. This was confirmed for the Solar System and statistically verified over a list of 443 exoplanets, using both present rotation periods and those estimated at the fiducial formation age of 650 Myr. The statistical probability to obtain these results by pure chance is $p<0.046$. Future measurements for more accurate values of stellar rotation are needed to reduce uncertainties and support the conclusions presented here.

The quantization in planetary orbits is a function of stellar rotation and consequently, is not related to any arbitrary universal constant but is specific to each exoplanetary system. Stellar rotation and the transfer of angular momentum play a key role in several planetary formation processes, such as tidal dissipation, disc truncation, and resonance, all of which could play a role in the resulting quantization reported here. Further investigation is required to understand the role of these physical processes.

The dependency on the central star's rotation rate corresponds to a strikingly similar relationship between the atomic orbital quantization (in integer multiples of the reduced Plank's constant $\hbar$) and the central proton's spin ($\frac{1}{2}\hbar$) in Bohr's model of the hydrogen atom. The quantization dependency on central body rotation, from atomic scales to large-scale gravitational systems, may be at the heart of a more general natural law or self-organizing principle that guides the formation of all rotating systems. Therefore, in addition to planetary systems, future investigation into the applications of this quantum-like model to other gravitational systems such as planetary satellites and ring systems, binary stars, and galactic centers, is intended to identify any discrete or quantized features in their orbital structure, and to validate the role and dependency of that quantization on the central body rotation rate. In concluding, the quantization of planetary orbits in half-integer multiples of the parent-star rotation periods is now a statistically significant fact that should not be ignored in future models of planet formation.

## 8. ACKNOWLEDGMENTS

We thank the anonymous referee for the valuable comments and for proposing the test in section 5.5.

This research has made use of the Exoplanet Orbit Database and the Exoplanet Data Explorer at http://exoplanets.org, as well as the Extrasolar Planets Encyclopedia at http://exoplanet.eu/ by Jean Schneider.